# Advances in electromagnetic techniques for subsurface infrastructure detection: A comprehensive review of methods, challenges, and innovations

Arasti Afrasiabi [1], Farough Rahimzadeh [2], Alireza Keshavarzi [3]

[1] Department of Civil Engineering, School of Engineering, University of Birmingham, UK,
Email: axa1213@alumni.bham.ac.uk
[2] Department of Civil Engineering, School of Engineering, University of Birmingham, UK
[3] Independent Researcher, Yazd, Iran

## 1 Abstract

This review paper explores the state-of-the-art in non-intrusive methods for detecting and characterising buried infrastructure, focusing on Electrical Resistivity Tomography (ERT), Infrared Thermography (IRT), and magnetometry, along with data fusion techniques and mathematical estimators. ERT and IRT offer distinct advantages in subsurface imaging, while magnetometry provides omnidirectional measurements ideal for detecting ferrous targets. Despite these benefits, each method has inherent limitations, such as challenges in depth estimation and difficulties in distinguishing between various subsurface objects. The integration of multiple sensing techniques through data fusion approaches has shown significant promise in overcoming these limitations and improving detection accuracy. Additionally, mathematical estimators, including Kalman filters and particle filters, play a crucial role in reducing noise and enhancing the precision of geophysical surveys. This review discusses the strengths, limitations, and future research needs of these techniques, offering a comprehensive understanding of their current and potential applications in buried infrastructure detection. The paper concludes by emphasising the importance of optimising sensor performance, refining fusion algorithms, and exploring hybrid models for real-time data processing in future research.

## 2 Introduction

The modern infrastructure of communities relies heavily on buried utilities for the distribution of vital resources such as water, electricity, sewage, gas, and telecommunication services. In the UK alone, there are over 4 million kilometres of buried infrastructure utilities (Bricker and Frith, 2017). However, like their above-ground counterparts, these underground assets are subject to ageing and deterioration, necessitating continuous maintenance, repair, and replacement to maintain serviceability (Davis et al., 2016; Samis et al., 2005). The complexity of modern urban infrastructure presents significant challenges in accessing and managing these buried utilities. Detecting and locating utilities is often complicated by their age, with many installed decades or even centuries ago (Rogers et al., 2012). Furthermore, the lack of well-documented maps providing accurate location information exacerbates this issue (Metje et al., 2012). Traditional ground investigation techniques, involving methods such as trenching and borehole drilling, are often costly, inaccurate, and potentially destructive to both the utilities and the surrounding ground (Makana et al., 2016). These methods can also lead to significant service disruptions and pose safety risks, with thousands of utility strike incidents reported annually in the UK alone (USAG, 2020, 2019).

In contrast, geophysical non-destructive testing (NDT) methods offer several advantages for subsurface investigation. These techniques can remotely investigate the subsurface by focusing on mechanical properties or by identifying contrasts in the physical properties of materials (Afrasiabi et al., 2025, 2023; Hao et al., 2012). NDT methods provide benefits such as rapid and accurate data gathering, cost efficiency, minimal disruption, and reduced risk of utility strikes, as well as attributes such as speed, non-contact surveying, ease of use, and portability (J Milsom and Eriksen, 2011; Pringle et al., 2021).

Among the various available geophysical NDT methods for shallow underground investigations, electromagnetic (EM) techniques—such as Ground Penetrating Radar (GPR), Electromagnetic Locator (EML), and Ground Conductivity Measurement (EMC)—have gained particular popularity due to their significantly faster data acquisition, higher detection resolution, and superior capability for detecting small or shallow anomalies, among other reasons (Ganiyu et al., 2020; Kalika et al., 2014; Zhou et al., 2019). Figure 1 summarises conventional EM geophysical techniques for detection and condition assessment of buried assets.

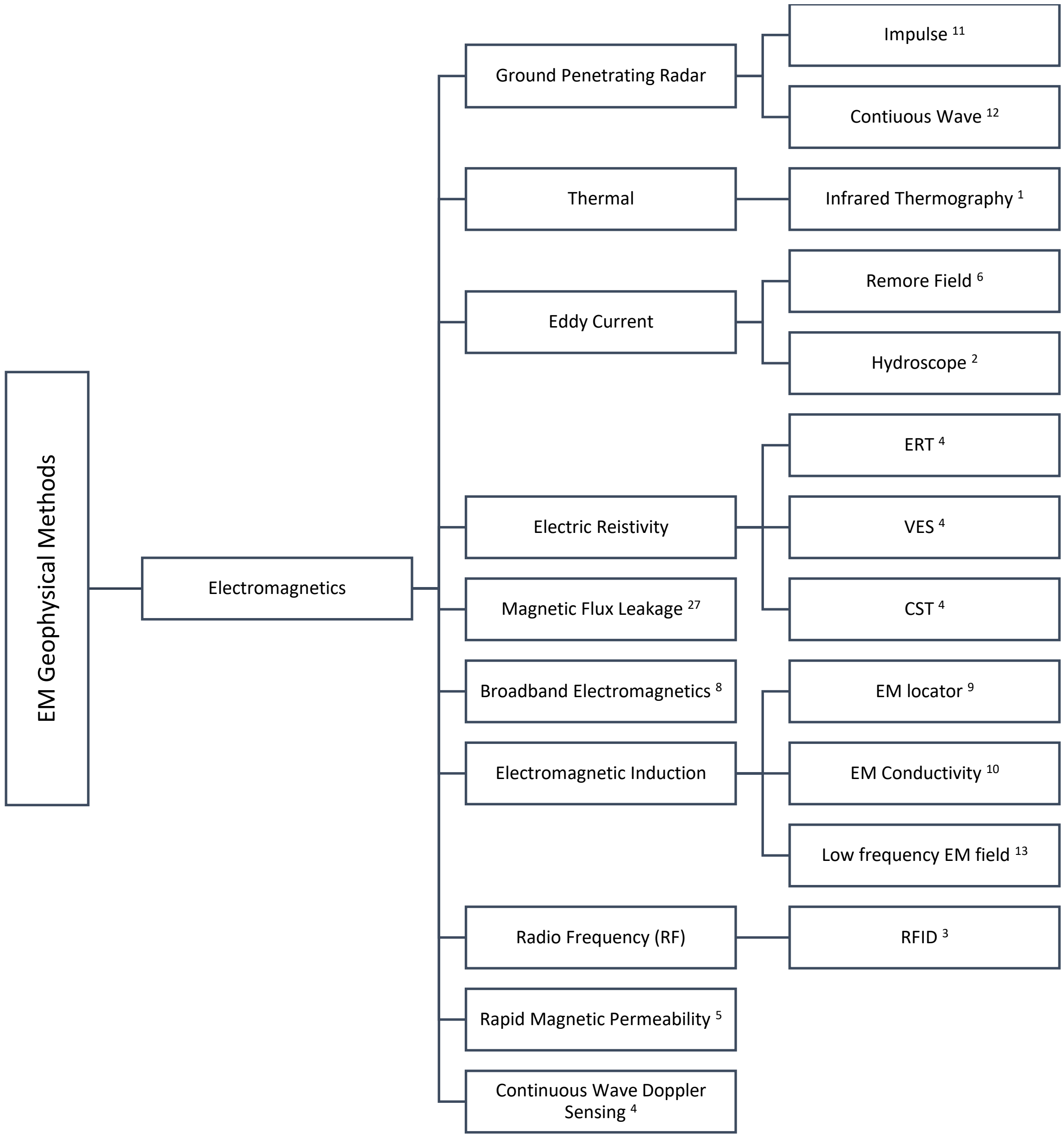

**Figure 1: Overview of classified non-destructive methods for detection and condition assessment[1]**

This paper provides a comprehensive review of electromagnetic (EM) techniques for subsurface utility detection and mapping, focusing on GPR, EML, and EMC as the most commonly used technologies in industry. It examines the capabilities, limitations, and potential synergies of these techniques, aiming to identify research gaps and highlight areas for future development in non-invasive subsurface infrastructure assessment.

---

1. (Carreño-Alvarado et al., 2014), 2. (Makar and Chagnon, 1999) , 3. (Abdelnour et al., 2018), 4. (Bimpas et al., 2010), 5. (Perrone et al., 2014), 6. (Perrone et al., 2014), 7. (Perrone et al., 2014), 8. (Roubal, 2018), 9. (Li, 2015), 10. (Jiles, 1990), 11. (Zheng and Liu, 2011), 12. (Siu and Lai, 2019), 13. (Moghadas, 2019), 14. (Pringle et al., 2020), 15. (Pasternak and Kaczmarek, 2019), 16. (Tabbagh and Panissod, 2000)

# 3 Ground Penetrating Radar (GPR)

## 3.1 Principles and Data Acquisition

Ground Penetrating Radar (GPR) is a widely used non-intrusive geophysical technique for imaging subsurface structures, including buried infrastructure. It offers easy data collection, high-resolution data, and real-time results (Daniels, 2004; Zajícová and Chuman, 2019). As an electromagnetic method, GPR operates by transmitting electromagnetic waves into the ground and measuring the reflected waves to obtain information about the subsurface (Daniels, 2004; Goodman and Piro, 2013a).

GPR systems utilise different antennas and a range of central frequencies (up to several GHz) to achieve varying penetration depths and resolutions. The central frequency refers to the dominant frequency in the spectrum of the emitted electromagnetic pulse, which is proportional to resolution and inversely proportional to penetration depth (Utsi, 2017). In other words, higher central frequencies provide better resolution but reduced penetration depth, while lower central frequencies allow for deeper penetration but with lower resolution (Benedetto et al., 2017). This flexibility allows GPR to be applied in various subsurface investigations, including: Identifying water tables in the ground (Bentley and Trenholm, 2002), mapping soil water content and moisture (Mahmoudzadeh Ardekani, 2013) differentiating layers and interfaces in ground composition (Binder et al., 2020; De Coster et al., 2018; Khamzin et al., 2017), detecting and locating buried infrastructure and service utilities (De Coster et al., 2019; Prego et al., 2017), water leakage detection (Bimpas et al., 2010; Lai et al., 2016), forensic investigations (Pringle et al., 2021, 2020, 2012), detecting and locating voids and cavities (Baek et al., 2018; Luo et al., 2019), heritage and archaeological explorations (Karamitrou et al., 2020), unexploded ordnance (UXO) detection (Abdelgwad and Nashat, 2017; Núñez-Nieto et al., 2015) concrete structure scanning and assessments (Sbartaï et al., 2012), road pavement inspection (Zhao and Al-Qadi, 2017), and asphalt crack detection (Liu et al., 2023).

The use of GPR is widespread across various industries, but interpreting GPR data is challenging and often requires expert analysis (Ciampoli et al., 2019). The complexity of GPR data necessitates skilled experts for accurate analysis and interpretation. Additionally, specialised software and techniques are required to effectively process and interpret the data (Wahab, 2013). The success of a GPR survey heavily relies on the contrast in electrical properties (impedance) of the subsurface materials. If this contrast is minimal or absent, the method may not be effective (Daniels, 2004). While GPR can identify anomalous targets, it cannot determine the composition of the material (Jol, 2009). These factors, within the resolution of penetration depth, can affect the performance of a GPR system in detecting anomalies under different circumstances (Wang et al., 2014). For example, a GPR system with a low central frequency may have difficulty detecting thin buried cables (Dou et al., 2017). The similarity of GPR frequencies with certain signal-generating equipment, such as mobile masts or medical equipment that produce electromagnetic fields, can create noise and mask the data (Reynolds, 2011). In addition, electromagnetic waves cannot propagate through highly conductive soils with lossy dielectric characteristics, which can cause rapid absorption of wave energy and result in shallow penetration of the GPR signal cables (Dou et al., 2017).

Conductivity, a measure of a material's ability to allow the flow of electric current, plays a crucial role in this limitation (Everett, 2013). High conductivity allows electrical current to flow easily, causing the waves to be absorbed rather than transmitted through the soil (Witten, 2014). Examples of lossy conductive soils include wet clays and salt-saturated soils (Farzamian et al., 2019). While many research papers have identified water content and saturation of soil as a source of noise and fuzzy areas in GPR radargrams (Bai and Sinfield, 2020; Bi et al., 2018; Ho et al., 2004; Solimene et al., 2014; Wai-Lok Lai et al., 2014), it is worth noting that water and soil saturation are not always a disadvantage for GPR and can be beneficial in certain circumstances. For example, water can improve the detectability of certain targets such as air-filled plastic or concrete pipes, which are normally difficult to detect with GPR due to their similar dielectric properties to the surrounding soil. The presence of water can make these targets more distinguishable by increasing the impedance between them and the surrounding soil (Bigman, 2012).

Strong detectable signals from high impedances, indicating contrast in dielectric properties, can suggest the presence of metallic objects, making them potential targets for GPR detection (Gonzalez-DIaz et al., 2020). However, GPR electromagnetic waves are unable to penetrate metals and will scatter upon interaction. This can lead to uncertainty when examining the ground beneath metallic objects such as steel plates, potentially rendering targets or soil layers invisible (Carrick Utsi, 2017a). Conversely, the corrosion of aged metallic objects, like cast-iron water mains, can heavily contaminate surrounding soil, resulting in highly conductive soil (Pennock et al., 2014). Studies on the corrosion of cast iron pipes and its impact on GPR detection by Mogharehabed (Mogharehabed, 2014), Schmidt et al. (Schmidt et al., 2006), and Pennock et al. (Pennock et al., 2014) indicate that the migration of iron ions into the soil can significantly alter the soil's dielectric properties. This alteration affects the propagation of GPR signals in terms of reflection, energy decay, and attenuation.

## 3.2 GPR Data characteristics

Sampled reflected GPR data can be visualised as A-scans which are waveforms of individual GPR traces (Figure 2a) consisting of peaks indicating reflectors with dielectric permittivity contrast (Skartados et al., 2019). Dielectric permittivity, a measure of a material's ability to store electrical energy in an electric field, affects the propagation and reflection of electromagnetic waves (Reynolds, 2011). The peaks along the trace are positioned based on the reflection time, which is also known as two-way travel time (TWTT) of the electromagnetic waves (Figure 2b). Plotting A-traces against distance creates radargrams (B-scan) which are radar cross-sections (RCS) created by plotting A-scans against distance (Figure 2c), providing indirect information about the subsurface structure in a scanned area perpendicular to the GPR moving direction (Benedetto and Benedetto, 2014). Signatures resembling hyperbolas in radargrams generated by concatenated A-scans can be used to locate buried features, with the position of the signatures indicating the actual position of the buried object underground (Benedetto and Benedetto, 2014).

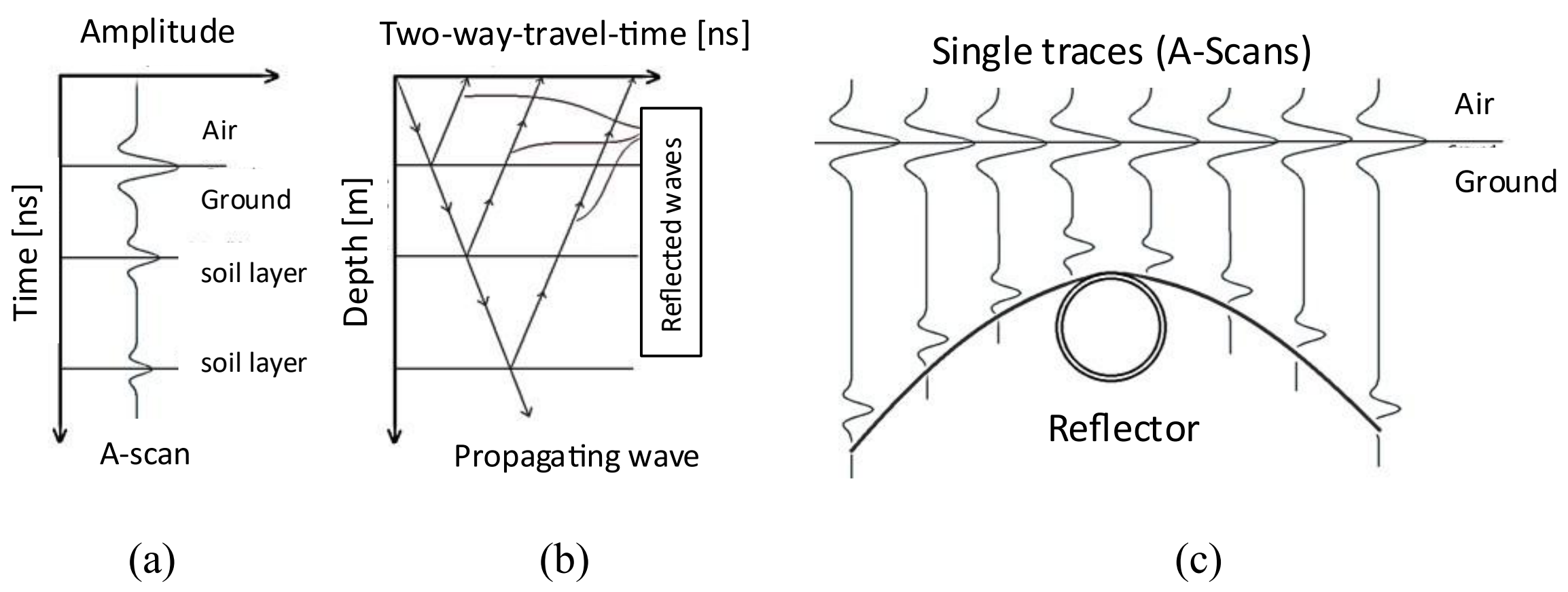

**Figure 2. schematic illustration of GPR data: a) A-scans, b) Two-way travel time, c) B-scans.**

## 3.3 GPR Data Processing

Signal processing plays a critical role in geophysical surveys, including GPR, by enhancing the clarity of data in both the time and frequency domains. It involves the use of computer-aided algorithms and filters to extract meaningful information and reduce noise from the collected data (Abraham, 2017; Benedetto et al., 2017; Van de Vegte, 2002). The success of a GPR survey depends on various factors, including the selection of an appropriate GPR device, equipment configuration, and proper survey conduct (Cassidy, 2009; Daniels, 2004). GPR data is often affected by noise and clutter, collectively known as background signals, which can obscure important information and impede data interpretation (Kumlu et al., 2020; Van Kempen and Sahli, 2001a). Dou et al. (2017) categorised noise sources in GPR data into three groups: system noise, non-homogeneity of the host medium, and interactions between transmitted and reflected signals.

To improve data interpretation quality, noisy data must be pre-processed and post-processed using signal and image processing techniques to reduce or eliminate noise and clutter (Allred et al., 2013). This process enhances data interpretability by highlighting meaningful signals and signatures in radargrams. A balance must be struck between removing unnecessary signals and retaining crucial target-associated information during processing (Utsi, 2017). This tailored approach is known as a supervised problem (Mallat, 2009). Common techniques for processing raw GPR data such as de-wowing, filtering, and signal gain adjustments, all aiming to enhance data quality and extract useful information (Che Ku Melor et al., 2021; Daniels, 2004; Economou et al., 2017; Rao et al., 2010). Iswandy et al. (2009) outlined a typical GPR data processing in three steps: define an appropriate framework, specify parameters for each planned step, and evaluate and improve results if necessary. The processing of GPR data typically involves several common basic approaches, which are outlined in Table 1.

**Table 1: Common basic approaches in processing GPR data**

| Method | Description | Reference |
|---|---|---|
| **Data Editing** | Applying, modifying, and rearranging information of data | (Park et al., 2019) |
| **Wave Velocity Estimation** | Estimating GPR wave propagation velocity through host medium | (Panda et al., 2022) (Greaves et al., 1996) |
| **De-Wowing** | Eliminating very low-frequency signals and DC bias components of the data | (Wong et al., 2016) (Jol, 2009) |
| **Time-Zero Correction** | Correction of start time to match with surface position | (Zadhoush, 2021) |
| **Signal gain adjustments** | Applying gains to the signal to improve the visibility and ease of interpretation | (Olhoeft, 2000) |
| **Horizontal Banding Removal** | Correction of data to ensure spatially uniform increments | (Downs and Jazayeri, 2021) |
| **Filtering** | One- and two-dimensions filtering to improve the signal to noise ratio and visual quality | (Kovalenko and Masalov, 2000) |
| **Deconvolution** | Contraction of signal wavelets to enhance reflection event | (Kadioglu, 2018) |
| **Elevation Correction** | Correcting for the effect of topography | (Busby and Merritt, 1999) |
| **Migration** | Corrections for the effect of survey geometry and spatial distribution | (Özdemir et al., 2014) |
| **Depth Conversion** | Conversion of two-way travel times (TWT) into depths | (Conyers and Lucius, 1996) |
| **Image Analysis** | Using pattern or feature recognition tools and methods | (Szymczyk and Szymczyk, 2013) |
| **Attribute Analysis** | Attributing signal parameters or functions to identifiable features | (Dossi et al., 2015) |
| **Modelling Analysis** | Simulating GPR responses making use of numerical models | (Warren et al., 2016) |

Although there are various techniques for processing GPR data (as shown in Table 1), four main techniques are most commonly used by practitioners and are considered the backbone of GPR data processing. These techniques address essential aspects such as timing and depth precision, signal amplification, noise reduction, and frequency enhancement (Huber and Hans, 2018). These techniques are as follows.

*Time Zero Correction*: This technique adjusts the start time of each trace to match the precise moment the radar pulse was transmitted. It provides a more accurate time reference, facilitating a more precise interpretation of the depth of subsurface features (Zadhoush, 2021).

*Signal Gain Adjustment*: As GPR signals attenuate with depth, this technique amplifies the scattered signals by multiplying the unprocessed initial trace (A-scans) by a gain function or coefficient. Linear and exponential gain functions, or a combination of both, are commonly used (Goodman and Piro, 2013b; Huber and Hans, 2018).

*Background Filter (Horizontal Banding Removal)*: This technique removes constant noise reflections that appear as horizontal banding in GPR radargrams. These can be caused by reflectors at a fixed distance from the antenna or by the GPR system itself (Downs and Jazayeri, 2021).

*Band-pass Filtering*: This method eliminates undesired signals outside the frequency range of a GPR system. It can be particularly useful for removing noise from external sources or amplified noise resulting from signal gain adjustments (Carrick Utsi, 2017b).

While these processing techniques can significantly improve data quality, it is crucial to apply them judiciously. Incorrect application can result in the loss of useful information or the introduction of artifacts (Goodman and Piro, 2013a).

## 3.4 GPR data Interpretation

Interpreting GPR data is inherently challenging due to the complexity of subsurface conditions and several factors that affect data quality—such as noise, structural complexity, limited depth penetration, incomplete datasets, and inhomogeneity of the subsurface (Dojack, 2012; Luo et al., 2023). Although signal and image processing techniques can help reduce noise and enhance clarity, they must be applied carefully to avoid removing meaningful information. A deep understanding of the limitations of GPR and the environmental factors influencing its performance is essential for achieving accurate and reliable interpretations. GPR data interpretation is often subjective, relying heavily on the expertise and experience of the interpreter. Both under-processing and over-processing of data can lead to misinterpretations: insufficient processing may leave noise or clutter that obscures key features, while excessive processing can distort or eliminate real subsurface signals. Effective interpretation, therefore, requires not only technical skill but also a clear understanding of the survey objectives and the limitations of the technology. Striking a balance between improving data quality and preserving authentic subsurface features is key to producing reliable and meaningful interpretations (Goodman and Piro, 2013b).

*Techniques for Hyperbola Detection in GPR Radargrams*

Manual techniques involve visual inspection, the use of cursors or measurement tools, manual curve drawing, and curve fitting. Although these approaches are time-consuming and susceptible to human error, they remain valuable—particularly for verifying the results of automated methods or analysing complex and ambiguous data where automation may fail.

Automated techniques, on the other hand, use algorithms to identify hyperbolic signatures more efficiently. These include methods such as the Hough Transform (Wang and Su, 2013),

Hyperbola fitting (Lei et al., 2019), and machine learning algorithms like neural networks (Amaral et al., 2022). While these techniques offer speed and consistency, they often demand considerable computational resources and may not generalise well across all GPR datasets due to variations in data quality, noise, and subsurface conditions.

Among these, the Hough Transform is one of the most widely used methods for detecting hyperbolas in GPR images (Yamaguchi and Mizutani, 2021). Capineri et al. (1998) introduced a modified version of the Hough Transform capable of detecting both hyperbolic and linear features in autonomous GPR B-scans. Building on this, Borgioli et al. (2008) combined the Hough Transform with time-of-flight analysis and the least squares algorithm to estimate the location of buried targets. Despite these advancements, several limitations have been identified. Zhou et al. (2019) highlighted challenges in detecting overlapping hyperbolas and accurately characterising their geometric properties. Similarly, Shihab and Al-Nuaimy et al. (2004) noted that the least squares fitting method may struggle with complex or noisy datasets, limiting its robustness.

Artificial Intelligence techniques like Artificial Neural Networks (ANN) and Genetic Algorithms (GA) have enhanced evaluation techniques. ANN has been successfully applied to GPR (Al-Nuaimy, 1999; Gamba and Lossani, 2000; Shihab et al., 2002; Szymczyk and Szymczyk, 2013; Travassos et al., 2018). Al-Nuaimy et al. (2000) and Shihab and Al-Nuaimy (2005) used neural networks with Hough transformation for hyperbola detection. Gamba and Belotti (2003) presented a neural network for buried pipe detection. Birkenfeld (2010) introduced an ANN approach for automated detection of healthy and interfered hyperbolas.

Recent advancements include machine learning-based frameworks for estimating rebar properties (Giannakis et al., 2020, 2019) and methods combining neural networks with image processing techniques (Dewantara and Parnadi, 2022). Amaral et al. (2022) conducted a systematic review of computer vision and machine learning algorithms for GPR image information extraction.

These methods show potential in automating buried object detection and property estimation using GPR, but also have limitations that should be considered. A wide range of methods have been employed by various researchers to detect patterns, fit curves, and process data in the context of image analysis, object detection, and signal interpretation. The Hough Transform was extensively used by Hart and Duda (1972), Capineri et al. (1998), Duda and Hart (2002), Yamaguchi and Mizutani (2021), Moeller et al. (2021), and combined with Least Squares methods by Borgioli et al. (2008). Cross-Correlation combined with Hough Transform was applied by Wang and Su (2013), while Ahn et al. (2001) used Least-Squares Orthogonal Distances Fitting, and Pilu et al. (2002) applied Direct Least-Squares Fitting. Various clustering approaches were used, including k-means clustering by Kanungo et al. (2002) and Cao et al. (2013), while Ertöz et al. (2013) used a general clustering method. Neural Networks (NN) have been a popular approach, used by Hajipour et al. (2023), Birkenfeld (2010), Gamba and Lossani (2000), Al-Nuaimy et al. (2000), Youn and Chen (2002), and Belotti et al. (2003), who combined NN with wavelet image processing. Faster R-CNN was employed by Lei et al. (2019), Zhang et al. (2021), and Srimuk et al. (2023), while Amaral et al. (2022) combined ANN and CNN. Evolutionary algorithms were also explored, with Genetic Optimizers used by Pasolli et al. (2009) and Multi-Objective Genetic Algorithms by Harkat et al. (2018). Afrasiabi (2023) and Afrasiabi et al. (2025, 2023) employed both Multi- and Single-Objective Genetic Algorithms, and Harkat et al. (2019) combined Neural Networks with Multi-Objective Genetic Algorithms. Chen (2010) introduced a Probabilistic Conic Mixture Model, Rosin (2001) applied Unimodal Thresholding, and Persico et al. (2010) used Microwave Tomography. Wavelet techniques were used by Idi and Kamarudin (2012a) via Digital Wavelet Transform, while Maas and Schmalzl (2013) applied the Viola-Jones Algorithm. Terrasse et al. (2015) utilized the Curvelet Transform, and Sagnard and Tarel (2016) applied Template Matching. Other fitting and transformation methods included Scale-Invariant Feature Transform (SIFT) by Harkat et al. (2016), Analytical Function Fitting by Mertens et al. (2016), Hyperbola Fitting by Rajiv et al. (2017), and Parabola Fitting with Retinography by García-Escudero et al. (2017). Bu et al. (2017) developed a Cluster-Aware Multiagent System, while Tzanis (2017) introduced a Curvelet-like Directional Filter, and Dou et al. (2017) proposed the Column-Connection Clustering (C3) Algorithm. Zhou et al. (2018) presented the Parabolic Fitting-Based Judgment (PFJ) and Open-Scan Clustering Algorithm (OSCA), while Luo and Lai (2020) used a Pyramid Method and Pattern Recognition. Ali et al. (2021) applied Hybrid Feature Extraction, and Li et al. (2020) implemented techniques using TensorFlow.

As discussed above, various techniques have been developed for specific GPR applications, each with limitations for cases outside their intended purpose such as edge detection issues in Hough Transform, overfitting in ANN, and high noise susceptibility in Clustering methods (Ali et al., 2021; Amaral et al., 2022; Ertöz et al., 2013; Manataki et al., 2021). The Hough Transform is discussed by several authors to have limitations such as noise sensitivity, computational intensity, and restriction to specific geometric shapes (Borgioli et al., 2008; Capineri et al., 1998; Hart and Duda, 1972; Wang and Su, 2013; Zhabitskaya et al., 2024). While robust to small deformations, it may struggle with larger ones (Borgioli et al., 2008; Zhabitskaya et al., 2024) and is threshold-dependent (Duda and Hart, 2002; Moeller et al., 2021). Least-squares methods, including orthogonal distances fitting (Ahn et al., 2001) and direct fitting (Pilu et al., 2002), may not always provide the best fit and are initial condition-dependent (Ahn et al., 2001). Clustering methods, such as k-means (Cao et al., 2013; Ertöz et al., 2013; Kanungo et al., 2002) and cluster-aware multi-agent systems (Bu et al., 2017), are sensitive to noise and initial conditions, and may struggle with overlapping clusters (Ertöz et al., 2013; Wunderlich et al., 2022). Neural networks, both ANNs and CNNs, have been employed by various researchers (Birkenfeld, 2010; Hajipour et al., 2023; Lei et al., 2019; Srimuk et al., 2023; Zhang et al., 2021). These methods require large datasets, are computationally intensive, and lack interpretability (Al-Nuaimy et al., 2000; Gamba and Lossani, 2000). Template matching (Sagnard and Tarel, 2016) is sensitive to noise and variations in scale, orientation, or appearance. The Faster R-CNN algorithm (Lei et al., 2019; Zhang et al., 2021) requires large datasets and may struggle with significant variations or deformations.

Other techniques used to detect hyperbolas in GPR data include the probabilistic conic mixture model (Chen, 2010), which has been found to potentially not work well with data that is highly nonlinear or has non-Gaussian noise (Chen, 2010), and unimodal thresholding (Rosin, 2001), which has been found to be sensitive to noise and potentially requiring pre-processing to smooth the image ( Ammar et al., 2023). Microwave tomography (Persico et al., 2010), digital wavelets transform (Idi and Kamarudin, 2012b), the Viola-Jones algorithm (Maas and Schmalzl, 2013), the curvelet transform (Terrasse et al., 2015), the Scale-Invariant Feature Transform (Harkat et al., 2016), fitting analytical functions (Mertens et al., 2016), the hyperbola fitting method (Rajiv et al., 2017), the parabolic fitting-based judgment (PFJ) technique (Zhou et al., 2018), the pyramid method (Luo and Lai, 2020), and TensorFlow (Li et al., 2020) have also been used to detect hyperbolas in GPR data, each with their own limitations as summarised in Table 2 below.

**Table 2. Summary of Hyperbola Detection techniques in GPR Radargrams; description, advantages, and limitations.**

| Method | Description | Advantages | Limitations |
|---|---|---|---|
| **Hough Transform** | Detects geometric shapes (e.g., hyperbolas) in GPR data. | Robust to partial and distorted features. | Sensitive to noise; computationally intensive; limited to simple shapes. |
| **Clustering Methods** | Groups similar data points into clusters to identify patterns. | Effective for large datasets. | Sensitive to noise, initial conditions, and overlapping clusters. |
| **Template Matching** | Compares GPR data to pre-defined hyperbola templates. | Works well for known shapes. | Struggles with variations in size, orientation, and appearance. |
| **Machine Learning (ML)** | Trains algorithms to identify hyperbolas based on features. | Adaptable to complex data. | Requires large training data; sensitive to noise; computationally expensive. |
| **Neural Networks** | Learns complex patterns directly from GPR data. | Handles non-linear patterns well. | Needs large datasets; lacks interpretability; computationally heavy. |
| **Curve Fitting** | Fits mathematical curves (e.g., hyperbolas) to data. | Effective for noisy, imperfect data. | Sensitive to noise; requires initial curve model; limited to simple functions. |
| **Cross-Correlation** | Measures similarity between signals to detect hyperbolas. | Simple implementation. | Limited to linear relationships; struggles with low-correlation data. |
| **Wavelet Image Processing** | Decomposes images into frequency bands to enhance features. | Enhances resolution and contrast. | Pre-processing required; computationally intensive. |
| **Probabilistic Conic Mixture Model** | Fits a mixture of conic shapes to data. | Models hyperbolas statistically. | Sensitive to initialisation and noise; requires good prior assumptions. |
| **Thresholding (Unimodal)** | Separates hyperbolas from background using intensity thresholds. | Simple and fast. | Works best on clean, unimodal data; noise-sensitive. |
| **Microwave Tomography** | Creates high-resolution subsurface images. | High spatial resolution. | Limited depth penetration; sensitive to subsurface conditions. |
| **Viola-Jones Algorithm** | Object detection trained on hyperbola images. | Effective if trained well. | Requires large, high-quality training data; struggles with variations. |
| **SIFT (Scale-Invariant Feature Transform)** | Detects distinctive features regardless of scale. | Robust to scale/rotation. | Not optimised for GPR data; computationally expensive. |
| **Column-Connection Clustering (C3)** | Clusters data points based on connectivity. | Detects connected hyperbola points. | Sensitive to noise; limited to small/medium datasets. |
| **Parabolic Fitting Judgment (PFJ)** | Fits parabolas to data to detect hyperbolas. | Simple curve fitting. | Sensitive to noise; needs good initial guesses. |
| **Pyramid Method** | Creates multi-resolution images to enhance features. | Enhances contrast/resolution. | Pre-processing needed; computationally intensive. |
| **TensorFlow-Based Detection** | Deep learning framework for hyperbola detection. | Highly adaptable. | Needs large datasets; sensitive to hyperparameters. |

As shown above, numerous approaches exist for manual and automated GPR data processing and hyperbola detection. This diversity stems from the complexity and variability of GPR data, with different approaches better suited for specific data types and applications. GPR data processing and hyperbola detection involve various trade-offs and challenges, including handling noise, errors, and uncertainties in the data, as well as selecting appropriate models or algorithms for data fitting. Consequently, researchers and practitioners have developed a range of approaches to address these challenges and optimise data processing and hyperbola detection tasks. However, no single approach can be classified as universally superior, as each method has its own strengths and limitations, performing differently depending on data characteristics and analysis goals. Some approaches may offer greater efficiency or accuracy, while others provide more robustness or flexibility. Furthermore, a method's performance may depend on its specific implementation or parameters and may vary with data size and quality. Therefore, it is crucial to carefully evaluate and compare different approaches to determine the most suitable one for a given application (Onyszko and Fryśkowska-Skibniewska, 2021; Solimene et al., 2014; Xiao et al., 2021; Youn and Chen, 2002).

However, GPR is not applicable in all site settings or for detecting all target objects. For instance, in the case of iron pipes, corroded material can attenuate the reflection and mislead monitoring practices (Pennock et al., 2010). Hao et al. (2012) concluded that GPR can be used to detect abnormally wet or abnormally dry areas in the ground, but in many cases, the water content of the soil leads to cloudy noise-polluted areas in the GPR image, obstructing the interpretation of results for detecting buried utilities. The detected voids and pipes in the literature using GPR are often in ideal conditions (e.g., minimal noise sources such as soil water content and uniformity of soil composition), which highlights the usefulness of GPR but neglects its limitations in more realistic scenarios where subsurface composition is complex and noise sources are influential. Despite these challenges, GPR remains widely used in industry, although its results are often dependent on expert interpretation, which can introduce subjective judgment, especially as the depth of the target increases and resolution decreases (Grandjean and Leparoux, 2016; Kofman et al., 2006; Xu et al., 2018).

The time-domain ultra-wideband (UWB) type of GPR has potential for more accurate location detection of erosion voids by widening the frequency range of the EM waves and acquiring more data compared to traditional GPR. However, this accuracy has only been reported for non-ferrous pipes in uniform dry soil (Jaganathan et al., 2010). Bimpas et al. (2011) reported that a GPR scanning antenna could detect a 40 cm diameter metal pipe buried at 1.5 m depth in controlled ideal settings. They concluded that their method had the potential to detect pipe leakage, although no practical evidence was provided.

As GPR has been in use for many years, various set-ups for its operation and interpretation have been suggested (Daniels, 2004). However, each has its own shortcomings, including limitations in detecting certain types of pipe materials, large attenuation in soils with high water content or salt minerals, and the crucial requirement of skilled operators for complex post-analysis, which is susceptible to subjective judgment (Chalikakis et al., 2011; Chen and Wimsatt, 2010; Hao et al., 2012; Metje et al., 2007). Aslam et al. (2018) reported that when there is no prior knowledge of ground wetness, GPR results can be prone to considerable uncertainty. Additionally, GPR survey data are often interpreted using image processing tools that identify hyperbolas in the GPR image, which represent anomalies such as buried pipes or cavities. Expert supervision and judgment in interpreting these results are crucial, but this also introduces subjectivity, which can limit the reliable application of GPR (Thitimakorn et al., 2016).

# 4 Electromagnetic Locator (EML)

## 4.1 Introduction to EMLs and their applications

The use of electromagnetic induction for geophysical surveys dates back to the nineteenth century, following Michel Faraday's groundbreaking discoveries in this field (Radiodetection, 2018). Electromagnetic Locators (EMLs) or pipe and cable locators (PCLs) have appeared as common non-intrusive methods use in the construction and utility detection sector for rapidly detecting metallic utilities in-situ (Bryakin and Bochkarev, 2019). In comparison to GPR, EMLs can only detect conductive masses (Chen and Huang, 1998). However, they are widely used in the UK due to their simplicity, low training requirements, small size, cost-effectiveness, and minimal data processing needs (Radiodetection, 2018). A comparison between GPR and EMLs is presented in Table 3.

**Table 3. Comparison of the relative advantages of EML and GPR**

| Subject | GPR | EML |
|---|---|---|
| **Equipment price** | High | Low |
| **Survey cost** | High | Low |
| **Equipment size** | Medium - Large | Small |
| **Equipment weight** | Light to heavy | Light |
| **Portable** | Yes | Yes |
| **Metallic utilities** | Yes | Yes |
| **Non-metallic utilities** | Yes | Yes*<br>*Only if equipped with tracing wire or<br>*Accessible for running the sonde into it |
| **Condition assessment** | Yes*<br>*Pipe leakage<br>*Pavement condition<br>*Rebars in concrete etc. | Yes*<br>*Some types of cable faults<br>*Coating condition |
| **Depth measurement** | Yes | Yes (under certain conditions**) |
| **Penetration depth** | Near-surface to deep | Near-surface |
| **Need expert** | Yes | Yes (needs simple training) |
| **All soil conditions** | No | Yes |
| **Underwater survey** | Yes (conditional) | Yes |

**** EMLs struggle with limited penetration depth due to variations in parameters such as emitted frequencies or electromagnetic properties of the ground** (Everett, 2013; John Milsom and Eriksen, 2011; Witten, 2014)**.**

## 4.2 Principles of Operation

Electromagnetic locators utilise quasi-static field (low-frequency field) method that uses time-varying electromagnetic fields (EMFs) to detect conductive masses in the subsurface (Huang et al., 2019). The basic components of these tools include:

*Signal emitting system*: Generates a primary time-varying electromagnetic field into the ground-conductor system.

*Search coils (magnetic sensors)*: Detect changes in the intensity of the Earth's magnetic field during the survey.

The EML is grounded in Maxwell's equations, particularly Faraday's Law of electromagnetic induction. When a time-varying electric current is applied to the transmitting coil, it generates a primary EMF that penetrates the ground (Witten, 2014). This primary EMF induces an alternating current (AC) in buried conductors, which in turn produces a secondary EMF around the conductor. This secondary EMF causes detectable changes in the Earth's magnetic field, which are measured by the tool's magnetic sensors (Wang et al., 2011). Moreover, the efficiency of AC signals over DC signals in generating detectable EMFs is explained by Faraday's Law. While DC signals create static magnetic fields that are difficult to detect, AC signals generate oscillating magnetic fields, resulting in higher voltages in the conductor and stronger, more easily detectable signals (Doolittle and Brevik, 2014; Huray, 2010).

In practice, buried conductors are located by measuring changes in the detected signals at different locations. The characteristics of the measured signal, such as its strength and position where it peaks, are used to determine the target's alignment, centre line, and potential depth (Fischer et al., 2002; Metje et al., 2007). The magnitude of the secondary EMF is proportional to the induced primary EMF, meaning that conductors further from the measurement point produce weaker responses. Typically, as illustrated in Figure 3, the maximum signal strength is measured directly above the conductor, with signal strength decreasing as distance from the conductor increases (Radiodetection, 2018). It's worth noting that AC currents in EML methodology can originate from both active sources (integrated into EMLs) and various other active and passive sources in the environment.

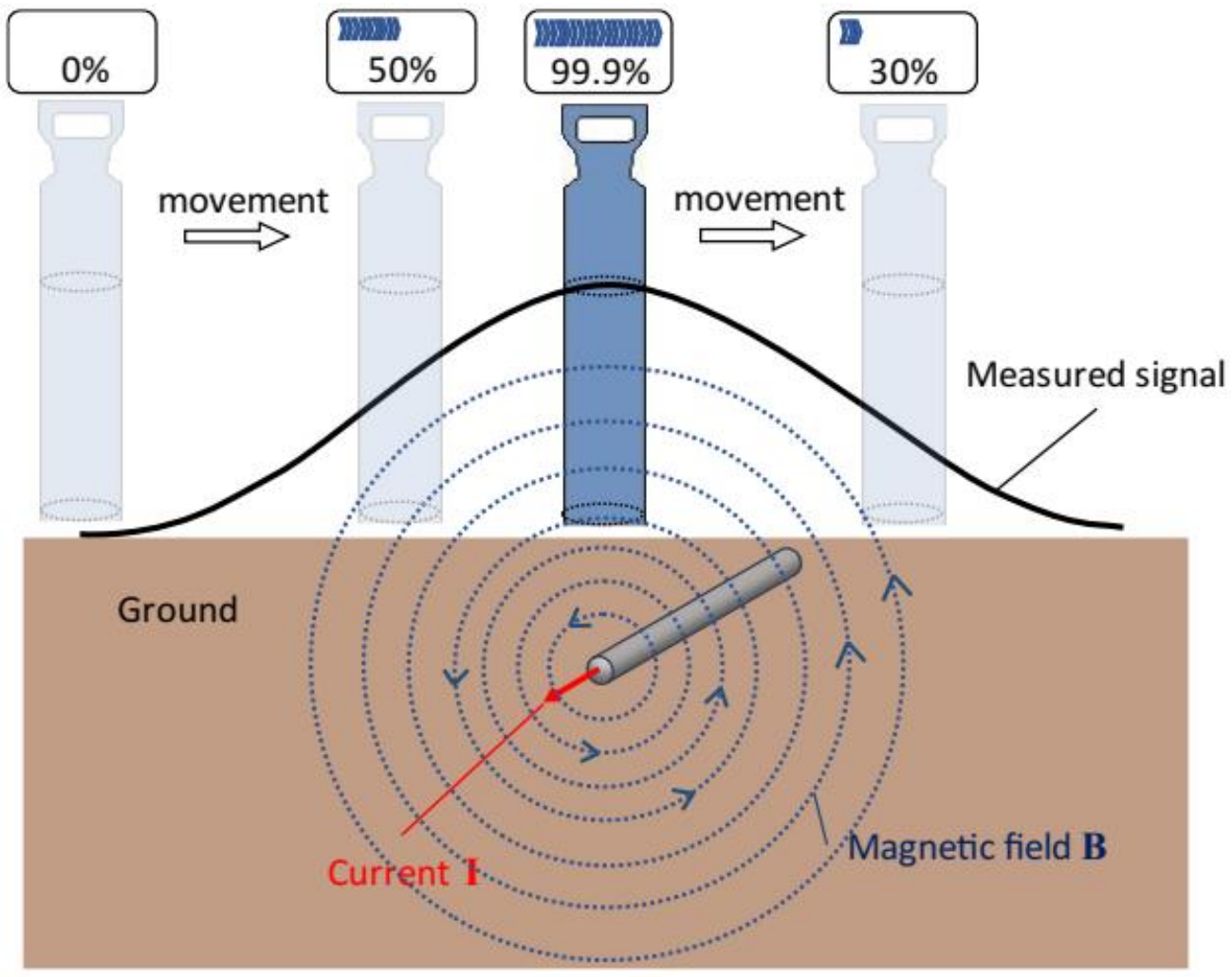

**Figure 3. Signal response from a conductor to EMF induction using cable avoidance tool (CAT) based on, After (Radiodetection, 2018).**

## 4.3 Active and Passive Detection Modes

EMLs can operate in both active and passive modes, providing flexibility in different survey scenarios:

*Active mode*

In this mode, the tool's integrated emitting system or an external compatible transmitter (often called a "Genny") generates a controlled electromagnetic field through an electrical current at specific frequencies (Bryakin and Bochkarev, 2019). The active mode allows for more precise control over the survey parameters and can be particularly useful when targeting specific types of utilities or when working in areas with multiple buried conductors. The current generated can be applied in three different approaches:

- Direct connection: The transmitter is directly connected to an accessible part of the conductive mass, such as a valve or meter (Figure 4a).
- Inductive clamping: A clamp is placed around an exposed cable or pipe to induce a current without direct electrical contact (Figure 4b).
- Inductive coupling: The transmitter is placed on the ground surface above the suspected location of the conductive mass, inducing a current in the conductor without direct contact (Figure 4c).

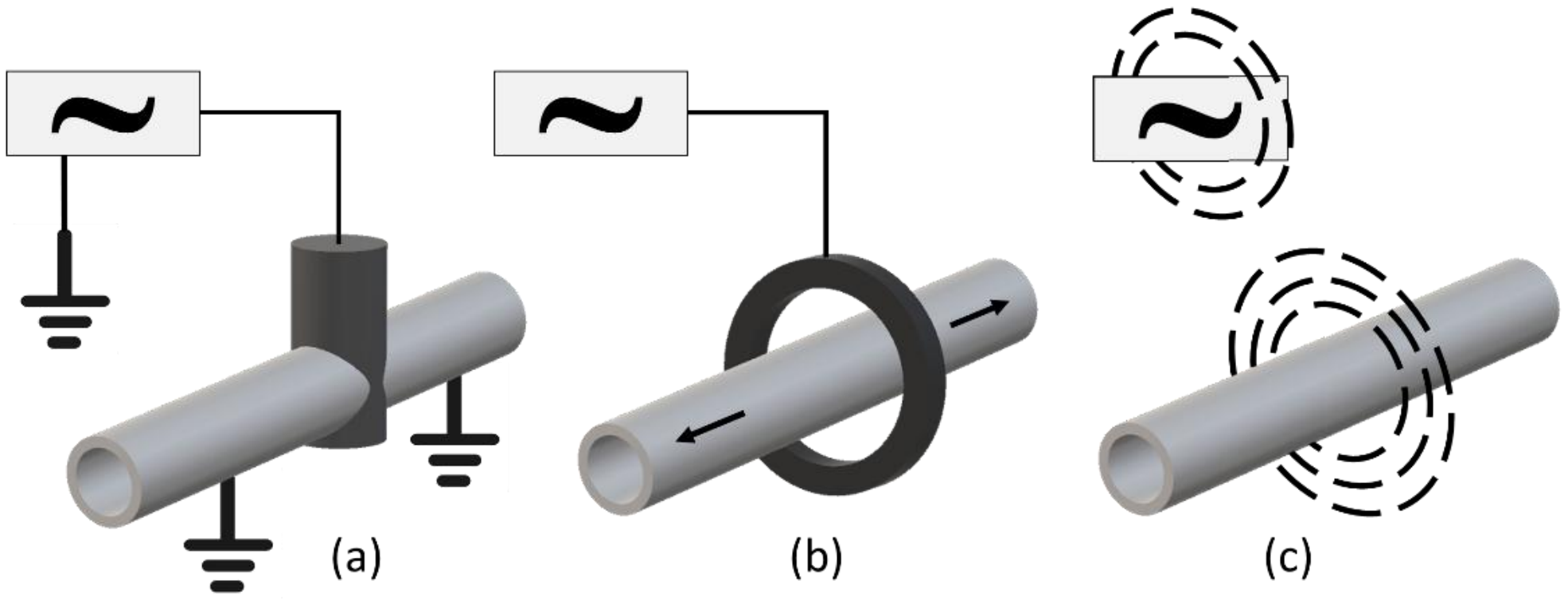

**Figure 4. a) Direct connection to induce AC, b) indirect connection by clamping the AC into the conductive mass, c) Inductive coupling without any direct access to the conductive mass.**

*Passive mode*

In this mode, the tool detects passive electromagnetic fields which naturally present in a conductive mass and can be detected by EMLs in their passive mode. This passive EMF generated by external sources, such as:

- Power mode: Detects the 50/60 Hz signals from live power cables.
- Radio mode: Detects very low frequency (VLF) radio signals that are reradiated by buried conductors.

The passive mode is useful for initial sweeps of an area and can detect utilities that may not be accessible for active signal application (Radiodetection, 2018; SanFilipo, 2000).

## 4.4 Depth Estimation

The strength of the secondary field is proportional to the induced primary field, meaning that conductive mass further from the measurement point will produce a relatively weak response, and vice versa (Figure 5). This principle allows operators depth of buried conductive mass based on the strength and characteristics of the measured signal. The simplest form of depth calculation is given by:

$$D = (d * A1) / (A2 - A1) \qquad \text{Eq. 1}$$

Where D is the estimated depth, d is the coil spacing, and A1 and A2 are measurements from the top and bottom coils, respectively. The accuracy of a simplified depth estimate using Eq. 1 may be affected by external distortion and noise sources such as adjacent utilities or other local obstructions above the ground. This issue can be addressed using the triangulation depth estimation method, which involves taking measurements from multiple locations to determine the buried conductor's depth accurately (Radiodetection, 2018; VIVAX, 2011; Waite and Wellaratna, 2006).

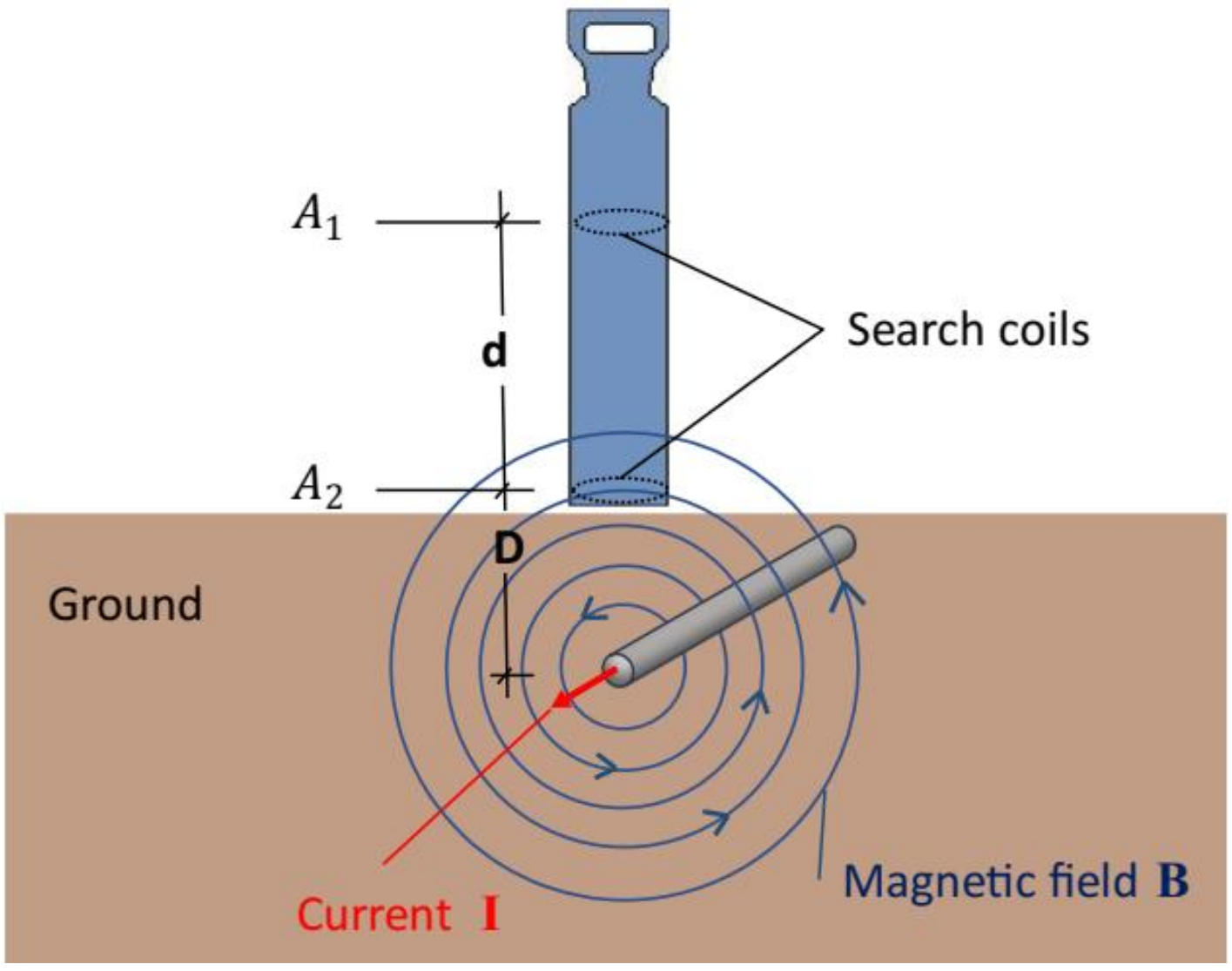

**Figure 5. The principle of utility detection using cable avoidance tool (CAT). it is important to properly connect the locator to the ground surface, as an incorrect connection will result in an inaccurate depth calculation, after (Radiodetection, 2018).**

## 4.5 EML method Limitations

While Electromagnetic Locators (EMLs) are widely used for subsurface utility detection, they are subject to several limitations that can affect their performance and reliability. These limitations stem from various factors including the soil conductivity, the size and conductivity of the conductor, the frequency and strength of the time-varying electromagnetic field (EMF) proportional to the AC currents, the grounding, and potential surrounding noise sources (Waite and Wellaratna, 2006). Understanding these limitations is crucial for practitioners and researchers to accurately interpret EML data and to drive further improvements in the technology. Table 44 summarises key limitations of the EML method, providing insights into current challenges and areas for potential future research and development in the field of subsurface utility detection.

**Table 4. Summary of Key Limitations and Challenges in Electromagnetic Locator (EML) Methodology**

| Limitation Category | Description | References |
|---|---|---|
| **Material Detectability** | Non-conductive utilities (e.g., concrete, plastic pipes, fiber-optic cables) cannot be detected unless a tracing wire is installed, or a sonde is used. | (Waite and Wellaratna, 2010)<br>(bin Haron et al., 2024) |
| **Soil Conditions** | Soil with high conductivity (due to high water or mineral content) causes faster AC decay in larger conductive masses, making them harder to detect or locate accurately. | (Jeong et al., 2003)<br>(Junoh et al., 2022)<br>(bin Haron et al., 2024) |
| **Signal Characteristics** | Higher frequencies: shallower penetration, rapid signal reduction.<br>Lower frequencies: deeper penetration, smaller reduction rate.<br>High frequencies can cause coupling effect, making it difficult to distinguish between different conductive targets close together.<br>In active EMF induction, a conductor can be detected if the frequency of the secondary EMF matches the frequency of the induced field. | (Akagi and Tamura, 2006)<br>(Dawalibi, 1986)<br>(Siu and Lai, 2019)<br>(Junoh et al., 2022) |
| **Data Recording and Processing** | Few older EML devices capable of recording data.<br>Some newer devices (e.g., Radiodetection RD8200) can log data, but processing capabilities are limited. | (Siu and Lai, 2019)<br>(Wang et al., 2010)<br>(Goddard et al., 2012)<br>(bin Haron et al., 2024) |
| **Multiple Utility Detection** | Difficulty in detecting multiple utilities in close proximity due to coupling effect and discrete measurements limitation.<br>Current approaches limited to areas with single power cables | (Siu and Lai, 2019)<br>(Wang et al., 2010)<br>(Goddard et al., 2012) |
| **Research Gaps** | 1) Limited published studies on logging and processing commercial CAT data<br>2) Lack of fusion algorithms for objectively processing and fusing EML data with other geophysical techniques (e.g., GPR) | |

### 4.6 Ground Conductivity Measurement (EMC)

Electromagnetic Conductivity (EMC), also known as EM induction or train conductivity, is an active electromagnetic technique used for measuring the conductivity of subsurface materials. While initially developed for natural resource (oil, minerals, etc.) exploration, it has found applications in detecting conductive materials and assets, as well as investigating ground lithography (Madden and Swift, 1965). Despite its low resolution, EMC offers a broader view of subsurface conditions compared to EML/CAT tools and can be particularly useful for mapping conductive utilities, detecting large-scale metallic structures (e.g., underground storage tanks), identifying areas of disturbed soil that may indicate buried infrastructure, and complementing other geophysical methods in integrated surveys (Cagniard, 1953; Madden and Swift, 1965; John Milsom and Eriksen, 2011; Moilanen, 2022; Stölben et al., 2022).

#### 4.6.1 Principles of Operation

EMC devices, or ground conductivity meters, consist of a controller and two separate wire loops (coils) serving as the transmitter and receiver. The transmitter coil generates a primary magnetic dipole field, which induces electrical currents in shallow depth conductors. These currents create a secondary magnetic dipole field measured by the receiver coil (John Milsom and Eriksen, 2011). The penetration depth of EMC is influenced by:

- Coil orientation (horizontal or vertical): Affects the depth of investigation and sensitivity to vertical or horizontal conductivity contrasts.
- Coil spacing: Larger spacing allows deeper penetration but reduces resolution.
- Frequency of the induced magnetic field: Higher frequencies provide better near-surface resolution but less depth penetration.
- Ground conductivity and composition: Highly conductive soils can limit depth penetration.
- Presence of metal objects: Can cause strong localised anomalies that may mask other features.

EMC devices typically operate at low frequencies (below 15 kHz) and measure apparent conductivity, which is a composite of ground conductivity responses to induced horizontal currents in horizontal ground layers (Beamish, 2011; McNeil, 1980; John Milsom and Eriksen, 2011; Petersen, 2001). These devices must be used under conditions of low induction number (*LIN*) in order to provide accurate conductivity measures. Based on this assumption, the ratio of the secondary magnetic field ($H_s$) to the induced primary magnetic field ($H_p$) is directly proportional to the ground conductivity ($\sigma$) of the subsurface, as shown in Eq. 2 (McNeil, 1980). This relationship allows for the rapid measurement of apparent conductivity without the need for ground contact.

$$\frac{H_s}{H_p} = \sigma \qquad \text{Eq. 2}$$

### 4.6.2 Data Interpretation and Inversion

Interpreting EMC data requires inversion techniques, which can be prone to non-uniqueness issues (Ku, 1976; Madden and Swift, 1965; Middleton and Valkenburg, 2002; Peyton, 2015). To simplify the inversion problem, it is common to convert the three-dimensional (3D) subsurface into one- or two-dimensional problems (Madden and Swift, 1965; Moilanen, 2022). It's crucial to note that geophysical problem inversions may not converge if the algorithm parameters are poorly chosen and initialised. Divergence issues can arise if the algorithm produces unrealistic outcomes, like negative conductivity values or negative soil layer thickness (Ku, 1976; McNeil, 1980; Stölben et al., 2022). Multiple researchers have proposed various implemented methodologies for the inversion of conductivity data in recent years and their methods can be categorised as presented chart in Figure 6.

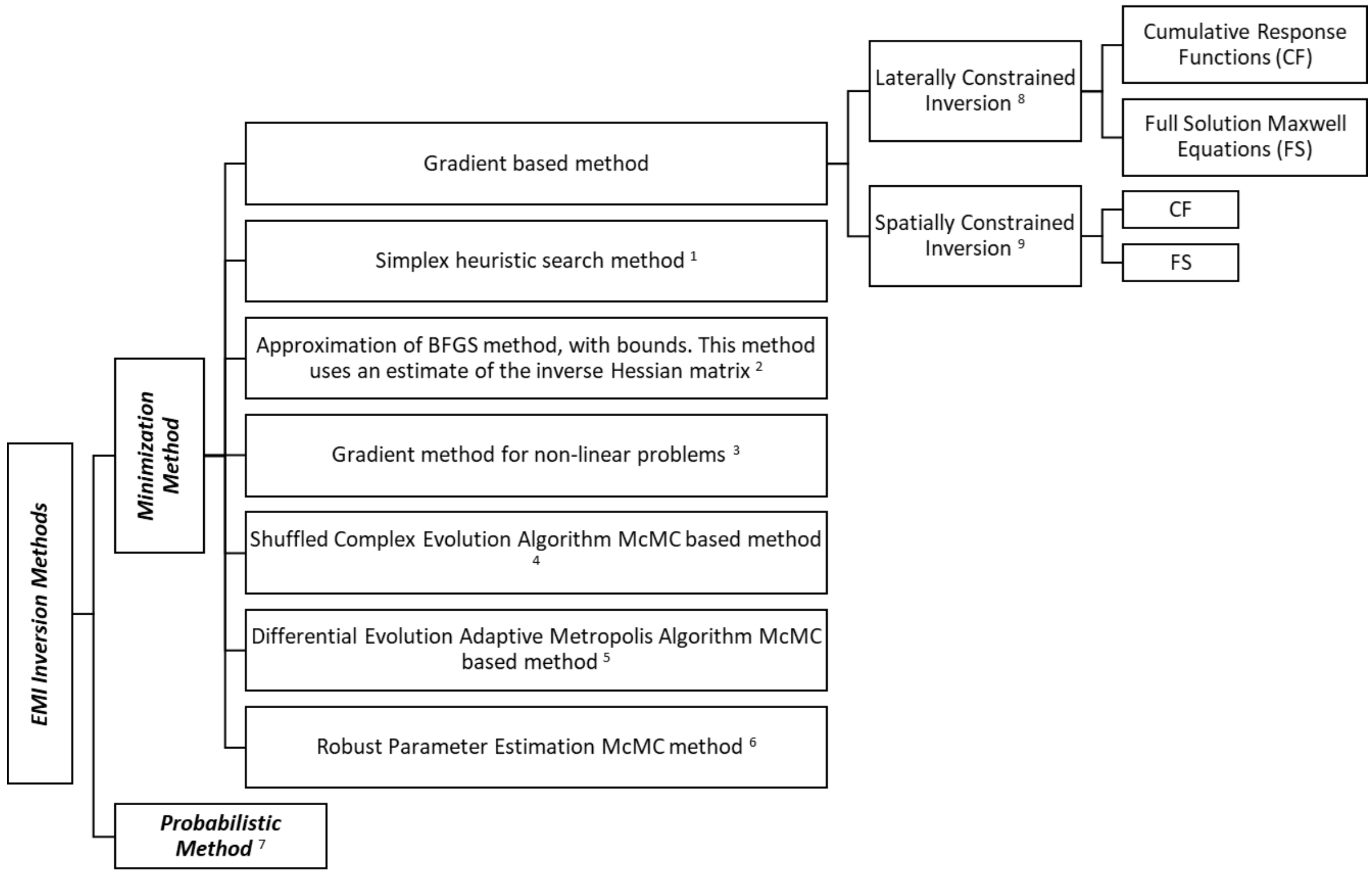

1. (Nelder and Mead, 1965), 2. (Byrd et al., 1995), 3. (Fletcher and Reeves, 1964), 4. (Duan et al., 1994), 5. (Vrugt and Ter Braak, 2011), 6. (Bárdossy and Singh, 2008), 7. (Moghadas, 2019), 8. (Farzamian et al., 2019), 9. (Monteiro Santos and El-Kaliouby, 2011).

**Figure 6. Inversion methodologies for inverting EMI data.**

Covering all of the methods presented in Figure 6 is beyond the scope of this article; however, regarding the usage of inversion algorithms for shallow-depth investigation, the following briefly discusses key studies that have utilised various inversion methods for EMC data. (1995)

Several studies have employed different inversion techniques to interpret electromagnetic conductivity (EMC) data in various geophysical applications. Sasaki (2000), and Sasaki (1989) utilised radio magnetotelluric and vertical magnetic dipole (VMD) devices for ground resistivity investigations, employing smoothness-constrained least-squares inversion methods combined with full solutions to derive conductivity profiles. In hydrogeological studies,

Monteiro Santos (2004) applied the EM34 device and used a modified 1D inversion with laterally constrained inversion and cumulative functions. Viezzoli et al. (2008) utilised the SkyTEM system for airborne time-domain electromagnetic surveys, employing spatially constrained inversion alongside 1D forward modelling to interpret the data.

For ground resistivity investigations, Christensen and Tølbøll (2009) applied radio magnetotelluric measurements at multiple frequencies and employed lateral parameter correlation inversion. In the context of geoelectrical structures, Santos et al. (2010) and Monteiro Santos et al. (2010) utilised the DUALEM-421 and DUALEM-21 devices, respectively, applying 1D inversion methods with laterally constrained inversion, sometimes accompanied by a full solution. For hydrostratigraphic feature analysis, Triantafilis and Santos (2011) combined the EM34 and EM38-MK2 systems with joint inversion, employing spatially constrained inversion and cumulative functions. Additional studies, such as those by Saey et al. (2015) and Melo and Régis (2017), explored resistivity imaging and geoelectrical structures using the DUALEM-21S and other devices, applying 1D laterally constrained smooth inversion and joint inversion methods, often alongside full solutions. Holtham et al. (2017) conducted airborne geophysical surveys with the SkyTEM system, applying smoothness-constrained inversion methods.

In soil science and hydrology, Guillemoteau et al. (2017) employed the DUALEM-21S device with 1D nonlinear inversion to analyze subsurface properties. Farzamian et al. (2019) focused on soil salinity, using the EM38-MK2 with laterally constrained inversion algorithms and full solution methods. Finally, Zare et al. (2020) utilized the DUALEM-421 device to map soil properties, applying specialized software (EM34soil) for inversion. These studies collectively demonstrate the diversity of approaches in EMC data interpretation, with inversion methods tailored to specific applications and technological advancements.

### 4.6.3 Limitations and Challenges of EMC Inversion Methods

In soil science and hydrology noise (Guillemoteau et al., 2017) employed the DUALEM-21S device with 1D nonlinear inversion to analyse subsurface properties. Farzamian et al. (2019) focused on soil salinity, using the EM38-MK2 with laterally constrained inversion algorithms and full solution methods. Finally, Zare et al. (2020) utilized the DUALEM-421 device to map soil properties, applying specialized software (EM34soil) for inversion. These studies, as well as other similar researches in this filed (Alipio et al., 2023; Guillemoteau et al., 2017; Lambot et al., 2009; Majewski et al., 2022; Moghadas, 2019; Monteiro Santos, 2004; Reynolds, 2011; Saey et al., 2015) collectively demonstrate the diversity of approaches in EMC data interpretation, with inversion methods tailored to specific applications and technological advancements.

### 4.6.4 Future Directions and Research Needs

Despite the widespread use of EMC surveys for detecting and locating conductive materials and assets in the subsurface, there is a lack of published studies combining EMC data with other remote sensing techniques. The integration of EMC with other geophysical methods can be improved, potentially leading to more accurate and comprehensive subsurface characterisation in various applications, including buried infrastructure detection.

# 5 Electrical Resistivity Tomography (ERT)

Electrical Resistivity Tomography (ERT) is a well-established, non-destructive method for subsurface exploration that utilizes the electrical properties of the ground and buried materials (Greggio et al., 2018; Kemna et al., 2002; Lesparre et al., 2019; Tso et al., 2017). The ERT technique operates by:

- Measuring electric potentials (voltages) between pairs of electrodes arranged in arrays on the ground surface using a high-impedance voltmeter.
- Interpreting these measurements as *apparent resistivity*, based on the assumption that the subsurface is homogeneous with a constant resistivity.
- Inverting the data to generate a 2D or 3D resistivity distribution map or cross-sectional image of the subsurface.

This inversion process estimates the parameters of a subsurface model that best correlates the observed electric potential values between electrode pairs with the actual resistivity distribution (Dey and Morrison, 1979; Galetti and Curtis, 2018; Pridmore et al., 1981).

The Electrical Resistivity Imaging (ERI) technique correlates distributed apparent resistivity with measured voltage by incorporating system parameters such as electrode separation, array configuration, and applied electrical current (Perrone et al., 2014). The penetration depth of the current increases with the spacing between electrodes, allowing the survey resolution to be adjusted according to specific objectives (Reynolds, 2011). ERT typically uses electrical currents that reverse direction at one- to two-second intervals, effectively functioning as direct current (DC) signals (John Milsom and Eriksen, 2011).

Although ERT requires electrodes to be inserted into the ground, it is still considered a non-intrusive method. This is because the survey depth is governed by the electrode spacing on the surface rather than the physical penetration depth of the electrodes (Kemna et al., 2002). The technique offers benefits such as cost-effectiveness and automated, software-controlled data acquisition. However, the manual task of electrode placement can be time-consuming and is often seen as a practical limitation (Kumar, 2012).

# 6 Infrared Thermography (IRT)

Infrared Thermography (IRT) is a non-invasive technique that has found applications from sustainability applications and energy efficiency of buildings (Al-Habaibeh et al., 2010) to medical purposes (Hildebrandt et al., 2010). Infrared waves are part of the invisible spectral region of the EM spectrum, which includes a full range of fluctuating electrical and EM waves passing through space (Herter, 2019; Zwinkels, 2015). According to Meola and Carlomagno (Meola and Carlomagno, 2004), IRT operates by:

- Capturing invisible infrared waves, which represent thermal energy.
- Transforming this thermal energy into visible images using infrared cameras.

In the context of buried infrastructure detection, Carreño-Alvarado et al. (Carreño-Alvarado et al., 2014) demonstrated the use of IRT with infrared cameras to locate underground pipes in urban areas, particularly during cold seasons. IRT leverages the temperature contrast between buried objects and the surrounding soil, making it easier to detect pipes when there is a significant difference between the pipe temperature and ambient ground temperature. Similarly, El Afi and Belattar (El Afi and Belattar, 2014)found that metallic pipes and those buried closer to the surface yielded better detectability using IRT. Going a step further, several studies have investigated the use of IRT to detect water main leakages beneath the ground surface conducted to detect and locate leakage in water mains beneath the ground surface using IR cameras (Atef et al., 2016; Fahmy and Moselhi, 2017; Shakmak and Al-Habaibeh, 2015). These studies reported successful leak detection with error margins ranging from 2.9% to 5.6%, particularly when the thermal contrast between the leaking pipe and surrounding soil was substantial. This suggests that IRT can be an effective tool for subsurface leakage detection under suitable environmental conditions.

The findings have shown that IRT is able to detect buried pipes under certain circumstances. However, its limitations with respect to the seasons and strong dependency on the ambient temperature (relative to the temperature of the pipe), soil type, and road construction can cause considerable limitations in utilising this method. Moreover, there are still gaps in knowledge in this field. One gap is the lack of clear understanding of the conditions under which the technique is likely to work. Another is the absence of a robust fusion algorithm for objectively processing and fusing IRT data with other geophysical methods for detecting subsurface assets.

# 7 Magnetometry

Magnetometry is a well-established non-intrusive technique that measures fluctuations in the Earth's magnetic field caused by ferromagnetic objects, such as metallic pipes and cables (Xie et al., 2013). It is also widely used for detecting unexploded ordnance (UXO) in ex-conflict or war regions (Zhang et al., 2003). In near-surface investigations, the Earth's natural magnetic field at the survey site provides a constant background field (MF), considered an external MF. Additional external MFs can originate from ferromagnetic objects above the surface, even if they are not directly over the measurement sensor. The fundamental challenge in magnetometry lies in distinguishing these external fields from those generated by subsurface objects (Everett, 2013; Robbes, 2006; Schlinger, 1990).

Magnetometers, the instruments used in these surveys, employ several sensing technologies. Fluxgate magnetometers detect magnetic flux leakage (MFL), while proton precession magnetometers (PPMs) measure the magnitude of the total magnetic field vector. Cesium vapour magnetometers, a type of atomic alkali vapour magnetometer, operate on a similar principle but utilise changes in the optical transparency of alkali vapours (Denisov et al., 2006; Robbes, 2006; Witten, 2014). Gradiometers, which consist of two vertically separated magnetometers, measure the magnetic field at slightly different depths, allowing calculation of the gradient to reduce external noise and enhance sensitivity to local subsurface anomalies.

Numerous studies have explored magnetometry for subsurface detection, especially for UXO detection and locating ferromagnetic pipes and cables. Zhang et al. (2003) and Billings (2004) developed signal processing and classification methods for magnetometry data, while Rodrigues et al. (2021) demonstrated the benefits of integrating magnetometry with complementary techniques, such as ground-penetrating radar (GPR). Numerical modelling using the Finite Element Method (FEM) has also been employed to analyse how object geometry, material properties, and burial conditions influence magnetic anomalies (Churchill et al., 2012; Guo et al., 2015).

Key findings from this body of work indicate that complex object shapes amplify magnetic dipole responses, and magnetic anomalies scale linearly with geomagnetic intensity, object thickness, and diameter. However, the orientation of objects introduces non-linear effects that complicate inversion and interpretation (Pan et al., 2016). Detecting thin and slender pipes remains particularly challenging due to their weak magnetic signatures (Soltanova et al., 2015). To improve performance, Feng et al. (2017)proposed the Magnetic Dipole Reconstruction (MDR) technique, which enhances depth estimation, segmentation, and computational efficiency.

Despite these advances, magnetometry suffers several limitations. It can only detect magnetic anomalies rather than directly identifying ferrous materials. Moreover, distinguishing between pipes, cables, and other ferrous objects is difficult, and accurate burial depth estimation remains challenging (Holt, 2014). Nevertheless, magnetometry offers the advantage of omnidirectional measurements, where targets do not need to lie directly beneath the sensor (Holt, 2014).

Another non-intrusive electromagnetic technique is Magnetic Flux Leakage (MFL), which requires close-range access to a buried object—such as a pipeline—to magnetise the object and scan the surrounding electromagnetic field using flux sensors (Bray and Stanley, 2018). Flaws in the object, such as wall breaks or cracks, alter the magnetic field and are detected in the MFL survey (Rizzo, 2010). MFL is primarily used for condition assessment of buried utilities at known locations where close-range access is possible. However, Mix (Mix, 2005) suggested that MFL could theoretically be extended to detect near-surface objects or assess surrounding soils. Significant challenges arise when adapting MFL for remote detection or condition assessment due to the complexity of subsurface environments and the interactions between buried objects and surrounding materials (Lai et al., 2018).

Other electromagnetic non-intrusive techniques for subsurface exploration include Remote Field Eddy Current (RFEC), Pulsed Eddy Current (PEC), and Ultra-Wideband (UWB) pulsed radar systems. These methods detect metallic or electrically conductive anomalies, such as cast-iron pipes or high-water-content zones resulting from leaks, by measuring the secondary magnetic fields produced by conductive buried objects within a primary induced field. However, the effectiveness of these techniques relies heavily on the electrical conductivity contrast between subsurface materials and requires prior knowledge of subsurface properties or buried utility locations for reliable interpretation (Liu and Kleiner, 2012). As a result, they share many of the same limitations as MFL and GPR.

The following table summarises the key advantages, limitations, and future research needs for magnetometry in shallow subsurface investigations.

**Table 5. Summary of advantages, limitations, and future research needs for magnetometry in shallow subsurface investigations.**

| Aspect | Key Points |
|---|---|
| **Advantages** | Omnidirectional measurements; targets need not lie directly below the sensor. |
| **Limitations** | Difficulty distinguishing between pipes, cables, and other ferrous objects; burial depth estimation challenges; cannot directly detect ferrous materials—only anomalies. |
| **Future Needs** | Validation under variable subsurface conditions; enhanced segmentation techniques; improved detection of complex and slender targets. |

In conclusion, while ERT, IRT, and magnetometry offer valuable capabilities for non-intrusive detection of buried infrastructure, each method has its strengths and limitations. The effectiveness of these techniques often depends on specific site conditions and the nature of the buried objects. As such, integrating multiple methods through data fusion approaches may provide more comprehensive and reliable results in detecting and mapping buried infrastructure.

# 8 Time-domain electromagnetic (pipe and cable locator) technique

The time-domain EM technique, also referred to as a pipe and cable locator, is an electromagnetic technique composed of a transmitter that emits radio frequency EM waves into the buried utility-ground system and a magnetic sensor that measures the magnetic field of the ground. The transmitter emits EM waves in the frequency range of 50 Hz to 480 kHz, depending on the intended penetration depth and the size of the object to be detected. Since the attenuation and penetration depth of the EM wave in the ground depend on the frequency of the EM wave and the ground conductivity, the higher the wave frequency, the shorter its penetration depth into the ground. As the emitted EM wave travels through the ground, it entangles with buried conductive objects, e.g. cast-iron pipes, in its projection (Metje et al., 2007). The entangled EM wave produces an altering electric current in the object proportional to the frequency of the EM wave. This current, in turn, produces an altering magnetic field that changes the ground's natural magnetic field. The Earth has a natural magnetic field due to the movement of ferrous liquid in the centre of the Earth. The alteration of this natural magnetic field, due to the local magnetic field induced by the EM wave, is measured with magnetic sensors on the surface and interpreted by skilled operators to infer the presence and location of buried conductive objects (Jeong et al., 2003).

In the case of buried pipes, the electric current induced in the pipe by this technique propagates along the pipe, producing a magnetic field along its length that enables tracing the pipe. The current attenuates as it propagates along the pipe, depending on the conductivity of the soil and the surface of the pipe, i.e. the size of the pipe. The current in larger pipes and in conductive soils, i.e. wet soil or soil with conductive mineral content, attenuates faster because the current scatters and escapes from the surface of the pipe into the conductive soil (Jeong et al., 2003). Furthermore, if the joints between the pipes in the pipeline are not conductive, e.g. rubber gasket joints, the current cannot propagate along the pipe. This reduces the induced magnetic field and, consequently, the possibility of detecting the buried pipe.

In general, the ground resistivity technique is suitable for detecting and locating buried metallic utilities, utilities that have a tracing wire installed above them, or when a part of the utility is accessible to directly induce the EM wave into the buried utility. However, for non-conductive buried utilities, e.g. concrete or plastic pipes and subsurface cavities, this technique cannot be used when there is no physical access to the inside of the pipe. In other cases, the successful application of this technique has been reported to depths of up to 2 metres (Sterling et al., 2009).

As an advantage, MFL can be used in passive mode to identify buried objects with a magnetic field. For example, for locating buried power cables, this technique can be used in passive mode, where the receiver measures the frequency of the altering magnetic field due to the altering current in the power cable without a transmitter. However, this sensitivity of MFL to buried power cables reduces its performance in identifying buried utility pipes when power cables are present in the vicinity (Kemna et al., 2002). Thus, the successful application of MFL requires sufficient a priori information to account for possible sources of error. There are reports that instruments with identical frequencies, antennas, and output signals were not equally able to detect the magnetic effect of the same buried object (Galetti and Curtis, 2018). This highlights the importance of appropriate settings depending on the target object, which requires a priori information about the ground and the buried target object. This includes information about adjacent conductive utilities, which can affect the magnetic field, e.g. a power cable close to a buried ferrous pipe reduces the possibility of detecting the pipe by altering the magnetic field.

Metal detectors, which work on a similar basis, are used to detect buried ferrous utilities. In metal detectors, a magnetic field is induced into the ground, which magnetises the buried conductive object. The magnetised object produces a secondary magnetic field around itself that can be measured from the surface. In conductive soils, however, this method cannot reveal the buried object because the induced magnetic field also magnetises the ground. Thus, the effect of the buried object cannot be measured (Jeong et al., 2003).

# 9 Frequency domain electromagnetic (Terrain conductivity) techniques

The Terrain Conductivity method is also known as the eddy current technique. In this technique, a time-varying magnetic field within the radio frequency range is induced into the ground via an EM transducer. This magnetic field generates an eddy current in the conductive material in the subsurface. An eddy current is a set of closed-loop electrical currents produced in conductive materials due to a magnetic field (Sophian et al., 2017). The eddy current in the conductive buried utility generates a secondary magnetic field that can be measured at the surface and interpreted to locate the buried utility. The Terrain Conductivity technique is a non-intrusive method that does not require access to the buried utilities and can be used to explore the subsurface to a depth of up to 3 m. However, this technique is sensitive to the conductivity of the soil. In other words, this technique is applicable where the conductive contrast between the buried utility and the surrounding soil is significant, e.g. a buried cast iron pipe in dry sand without conductive mineral content. Otherwise, detection of cavities in the ground cannot be reliably performed using this technique (Sophian et al., 2017).

# 10 Data Fusion and Multi-Sensor Integration

In addition to traditional data processing and interpretation techniques, data fusion provides complementary approaches for non-intrusive survey methods used in detecting buried infrastructure. It enables researchers to overcome the limitations of individual sensors and enhance data quality and interpretation accuracy by combining multiple data sources (Gros, 1997a). To achieve this, the data collected from multi-sensor surveys must be processed, fused, and interpreted systematically and objectively (Samadzadegan et al., 2025). This section explores the different levels of data fusion and their applications in detecting buried infrastructure.

## 10.1 Levels of Data Fusion

It is essential to use the most appropriate fusion method for the survey to enhance accuracy and minimize errors (Gros, 1997b; Liggins et al., 2009; Samadzadegan et al., 2025). The participating sensors must be combined using a distribution scheme to facilitate communication and data exchange within a chosen fusion algorithm (Yang et al., 2019). This refers to the order and arrangement of sensors, which can be divided into two main categories: parallel and serial configurations, and a combination of these as a subcategory. These categories are based on the types of sensors, survey configuration, and boundaries (Gros, 1997a; Oliveira et al., 2022; Qiu et al., 2019). NDT Data fusion can be categorised into three main levels: *low, intermediate, and high* (Faundez-Zanuy, 2009; Gros, 1997a; Oliveira et al., 2022). Each level serves a specific purpose in the data integration process and offers unique advantages for buried infrastructure detection.

Data fusion in subsurface detection can be categorized into three main levels—low, intermediate, and high—each offering distinct advantages depending on the application and sensor configuration. Low-level fusion combines raw data from multiple sensors to create a unified dataset, often improving signal quality and feature extraction. Techniques such as multiresolution analysis (MRA) allow for better visualisation of subsurface structures, especially in applications like ground-penetrating radar (GPR), while optimisation methods, including genetic algorithms, help improve the integration of various data sources. Heuristic methods, such as artificial neural networks, can also process and combine raw data to enhance feature recognition.

Intermediate-level fusion focuses on integrating extracted features from different sensor data sources to generate thematic maps for more accurate interpretation and detection. Methods such as neural networks are employed for pattern recognition and classification, while clustering algorithms help group similar features across different data sources. High-level fusion, on the other hand, uses statistical and stochastic techniques to combine sensor data, providing outputs with enhanced reliability. Approaches like Bayesian theory and the Dempster-Shafer paradigm are used to assess probabilities and combine evidence from diverse sensors, allowing for more robust decision-making in complex environments. Each fusion level, from low to high, plays a critical role in refining data interpretation and improving the detection of buried infrastructure.

In the context of buried infrastructure detection, the choice of fusion level depends on the specific sensors used, the type of infrastructure being detected, and the environmental conditions. Low-level fusion might be preferred when dealing with similar sensor types (e.g., multi-frequency GPR), while high-level fusion could be more appropriate when combining fundamentally different sensing modalities (e.g., GPR and electromagnetic locators). Table 6 presents a list of reviewed works in this area.

**Table 6. List of works focused on remote sensing data fusion**

| Author | Devices | Method | Purpose | Fusion Level |
|---|---|---|---|---|
| (Steinway et al., 1998) | GPR(SFCW) + Infrared + Metal Detection | Logic joint Fusion | Landmine detection | High |
| (Weisenseel et al., 1999) | GPR + EMIS[2] | Optimal maximum likelihood Method | Detection of plastic A/P mines | High |
| (Collins et al., 2002) | GPR + EMC2[3] | Bayesian detection approach | Landmine detection | High |
| (Ligthart and Yarovoy, 2003) | GPR + GPR (SFCW[4]) | Inverse scattering method | Landmine detection | Low |
| (Scott et al., 2004) | GPR + EMC + Seismic | Logic joint Fusion | Landmine detection | High |
| (Moghadas et al., 2010) | GPR + EMC | Comparison of various fusion methods: Integrated full-waveform EMC and GPR inversion, | Estimating soil electrical properties | Low-High |

[2] Electro-Magnetic Induction Spectroscopy
[3] Electromagnetic Conductivity
[4] Stepped-Frequency Continuous Wave

| | | | | |
|---|---|---|---|---|
| | | Weighting methods, Bayesian data fusion, Sequential inversion | | |
| (Lim and Cao, 2013) | GPR + IR[5] + UPV[6] | Logic joint Fusion | Sidewalk and concrete stairwell | High |
| (Dutta et al., 2013) | GPR + Vibro acoustic | Dynamic Bayesian network | Buried plastic water pipe | Intermediate - High |
| (Li et al., 2015) | GPR + GPS[7] + GIS[8] | Hybrid GPR/GPS + 3D probabilistic uncertainty band | Mapping and visualisation of underground utilities | Low-High |
| (Dou et al., 2017) | GPR + Vibro acoustic + PMF[9] + MG[10] + LFEM[11] | Marching-Cross Section (MCS) algorithm | Locating buried utility segments | Low - Intermediate |
| (Capozzoli and Rizzo, 2017) | GPR + infrared + ERT[12] | Logic joint Fusion | Concrete Structure | High |
| (Knox et al., 2017) | GPR + EMI[13] | Logistic regression, Gradient boosting decision trees | Buried Explosive Threat Detection | Low - Intermediate |
| (Smith et al., 2017) | GPR + WEMI[14] | Aggregation of different Choquet integrals within and across sensors | Explosive hazard detection | High |
| (Hafsi et al., 2017) | GPR + EML (no data) | Merge using a linear combination | Detection and localisation of underground networks | Intermediate - High |

[5] Impulse Response
[6] Ultrasonic Pulse Velocity
[7] Global Positioning System
[8] Geographical Information System
[9] Passive Magnetic Fields
[10] Magnetic Gradiometer
[11] Low Frequency Electromagnetic Fields
[12] Electrical Resistivity Tomography
[13] Electro-Magnetic Induction
[14] Wideband Electro-Magnetic Induction

| | | | | |
|---|---|---|---|---|
| (Xu et al., 2018) | GPR + GPR (Multi Frequency) | Fourier transform | Underground Disease Detection | Low |
| (Zhou et al., 2019) | GPR + Electric field device | Electric Field Locating + B-scan image interpreting | Plastic pipe detecting | Low - Intermediate |
| (Li et al., 2019) | GPR + Camera | 3D reconstruction method | Subsurface Pipeline Mapping | Intermediate |
| (Bianchini Ciampoli et al., 2020) | GPR + InSAR[15] | Data fusion logic | Assessment of linear transport infrastructures | High |
| (Lu et al., 2020) | GPR + GPR (Multi Frequency) | Image fusion based on 2D wavelet transform | Multi-frequency and multi attribute GPR | Intermediate |
| (Adamopoulos et al., 2021) | GPR + Laser scanning + imaging + thermography | Data fusion logic | Built heritage survey | High |
| (Massaro et al., 2021) | GPR + infrared | Long term short memory neural network algorithm | Road infrastructure inspection | High |

This section introduces the use of mathematical estimators such as Kalman filters and particle filters, which are commonly employed in various fields like autopilots, navigation systems, and self-driving cars, to achieve data fusion and state estimation. The primary purpose of using these estimators in subsurface mapping and exploration is to mitigate noise and uncertainty, particularly in GPR (Ground Penetrating Radar) data and other geophysical measurements. By automating the processing and integration of data from multiple sources, these estimators can significantly enhance the accuracy and reliability of subsurface maps.

The use of these advanced mathematical tools allows for the effective fusion of multi-sensor data, such as from GPR and other geophysical methods, which in turn improves the final output. This reduces the reliance on manual interpretation, thus making the detection of buried infrastructure more efficient and precise (Afrasiabi et al., 2025, 2023). However, the key challenge is the careful selection and fine-tuning of the appropriate estimator for each specific case, as various estimators may exhibit different strengths and weaknesses depending on the characteristics and objectives of the data involved. In the next section, a critical review of the mathematical estimators used for data fusion in GPR applications will be presented.

[15] Interferometric synthetic aperture radar

## 10.2 Mathematical Estimators for Data Fusion

Data fusion techniques play a crucial role in integrating information from multiple non-intrusive sensing methods to improve the detection and characterisation of buried infrastructure. Mathematical estimators can be broadly categorised into two types (Bar-Shalom et al., 2001; Särkkä, 2013; Simon, 2006):

*Stochastic estimators*: Based on probabilistic models that incorporate uncertainty (e.g., Kalman Filter (KF), Markov chain Monte Carlo (MCMC)) that update predictions of a system's state based on sensory data and a system dynamics model.

*Statistical estimators*: Based on statistical models of data distributions (e.g., Maximum Likelihood Estimator, Bayesian estimators) that estimate the parameters of a statistical model and are often used to estimate subsurface properties from GPR data.

While they can significantly improve the accuracy and reliability of data processing, they also present challenges, including computational resource requirements and the risk of model misspecification or overfitting (Särkkä, 2013). Table 77 presents mathematical estimators that have been used in GPR applications.

**Table 7. Mathematical estimators used in GPR applications**

| Method | Description | Reference |
|---|---|---|
| **Linear Kalman filter (LKF)** | A recursive estimator that updates a prediction of the system's state based on sensory data and a model of the system's dynamics | (Simon, 2006)<br>(Carevic, 1999)<br>(Afrasiabi et al., 2023)<br>(Afrasiabi et al., 2025) |
| **Extended Kalman filter (EKF)** | A variant of the Kalman filter used to estimate the state of a nonlinear model. Linearises the model around the current estimate of the parameters and applies the Kalman filter to the linearised model. | (Särkkä, 2013)<br>(Eppstein and Dougherty, 1998) |
| **Unscented Kalman filter (UKF)** | A variant of the Kalman filter that uses deterministic sampling to capture the mean and covariance of the state distribution. It achieves higher accuracy than the EKF by propagating carefully chosen sample points through the true nonlinear model, avoiding the need for linearisation. | (Särkkä, 2013)<br>(Nguyen and Pyun, 2015) |
| **Markov Chain Monte Carlo (MCMC)** | A method that uses random sampling from a probability distribution, where each sample depends only on the previous one, to estimate the parameters of a statistical model. | (Gallagher et al., 2009)<br>(Wang et al., 2021) |
| **Particle filter** | A Monte Carlo method that uses a set of particles to represent the distribution of the model's parameters. | (Ng et al., 2008; Simon, 2006) |
| **Interacting multiple model (IMM) filter** | A method that combines multiple Kalman filters to handle systems with multiple operating modes or models. | (Mazor et al., 1998) |
| **Maximum Likelihood Estimator (MLE)** | A statistical estimator that estimates the parameters of a statistical model by maximising the likelihood function. | (Myung, 2003; Nadim, 2008) |
| **Bayesian estimator** | A statistical estimator that uses Bayesian inference to estimate the parameters of a statistical model. Combines the observed data with a prior distribution over the parameters to form a posterior distribution, which is used to estimate the parameters. | (Luo and Fang, 2005; Särkkä, 2013) |

In GPR applications, various mathematical estimators are used to enhance data processing for buried infrastructure detection. Below is a summary of the advantages and limitations of these methods extracted from the publications above.

*Linear Kalman Filter (LKF)*

The LKF is computationally efficient and provides optimal estimates for linear systems with Gaussian noise. Its recursive nature allows for real-time processing, making it suitable for dynamic applications such as GPR. However, its primary limitation is that it is restricted to linear models, making it unsuitable for non-linear systems. The accuracy of the LKF is highly dependent on the initial estimates, and if these estimates are far from the true values, convergence may be difficult. Additionally, the LKF assumes Gaussian noise, which may not always represent real-world conditions, potentially leading to errors in the estimates.

*Extended Kalman Filter (EKF)*

The EKF extends the Kalman Filter to non-linear systems by linearising the system model around the current estimate. This makes it more versatile than the LKF, as it can handle non-linear models. It is still computationally efficient and suitable for real-time applications. However, the EKF shares many of the same limitations as the LKF, such as sensitivity to initial estimates and a reliance on Gaussian noise. Furthermore, the process of linearising a non-linear model may introduce errors if the linear approximation is not accurate.

*Unscented Kalman Filter (UKF)*

The UKF improves upon the EKF by using a set of carefully selected sample points to represent the state distribution, which avoids the need for linearisation. This makes it particularly effective for non-linear models, providing more accurate estimates even when significant non-linearities are present. However, the UKF is more complex to implement than the LKF or EKF, and it remains sensitive to model parameters, which can affect its accuracy if not tuned properly.

*Markov Chain Monte Carlo (MCMC)*

MCMC is a powerful method for handling complex models with unknown or non-standard probability distributions. It provides a probabilistic representation of the parameter estimates, making it useful for systems with inherent uncertainty. However, MCMC is computationally intensive and requires a large number of iterations to converge, especially for high-dimensional systems. It also requires that the probability distributions of the system are known or can be approximated accurately, which may not always be feasible.

*Particle Filter*

The particle filter is a Monte Carlo method that can handle non-linear and non-Gaussian models. It represents the probability distribution of the system's parameters using a set of particles, making it particularly useful for dynamic systems where the model's structure is not easily captured by linear or Gaussian assumptions. However, particle filters can be computationally expensive, especially for complex models, as they require maintaining a large number of particles. They are also sensitive to model errors and the choice of initial estimates, which can significantly affect the accuracy of the results.

*Interacting Multiple Model (IMM) Filter*

The IMM filter is used when a system has multiple modes of operation, combining different Kalman filters to handle each mode. It can provide a probabilistic representation of the system's parameters, improving accuracy by leveraging information from multiple models. However, the IMM filter is more complex to implement than a single Kalman filter, as it requires a switching mechanism to determine which model is most appropriate at any given time. It also shares the sensitivity to model parameters and requires accurate model switching to avoid errors in the estimates.

*Maximum Likelihood Estimator (MLE)*

MLE is a statistical estimator that maximizes the likelihood of observing the given data, making it robust to noise and outliers. It is particularly useful in situations where the underlying model is complex or uncertain. However, MLE is computationally expensive and requires good initial estimates. Like the other methods, it can be sensitive to the quality of the initial estimates, and if the starting parameters are far from the true values, the estimates may be inaccurate.

*Bayesian Estimator*

The Bayesian estimator incorporates prior knowledge or expert knowledge into the estimation process by combining it with observed data to form a posterior distribution. This can be beneficial for systems with limited or noisy data, as the prior information helps guide the estimation. However, the method requires a specified prior distribution, which may be difficult to define accurately, and its performance depends heavily on the choice of the prior. Additionally, like other probabilistic methods, it is computationally intensive and sensitive to initial estimates.

## 10.3 Kalman Filter and Its Case Studies

KF was first introduced by Rudolf Emil Kalman in 1960 as a recursive algorithm for probabilistic parameter estimation. It uses a system model, defined by a set of ordinary differential equations, to predict the behaviour of a model over time, based on sensor measurements (Brunton and Kutz, 2017; Simon, 2006); however, it is well-known that the system model and sensor measurements in a survey are not always accurate and reliable. Therefore, in KF the level of confidence in both system's model and sensor measurements should be determined by the experts. These parameters will be used in KF to calculate the Kalman gain and level of confidence of estimations at each time frame after each measurement. With having the Kalman gain, update its prediction and generate new estimates, Figure 7 shows the KF process in a diagram.

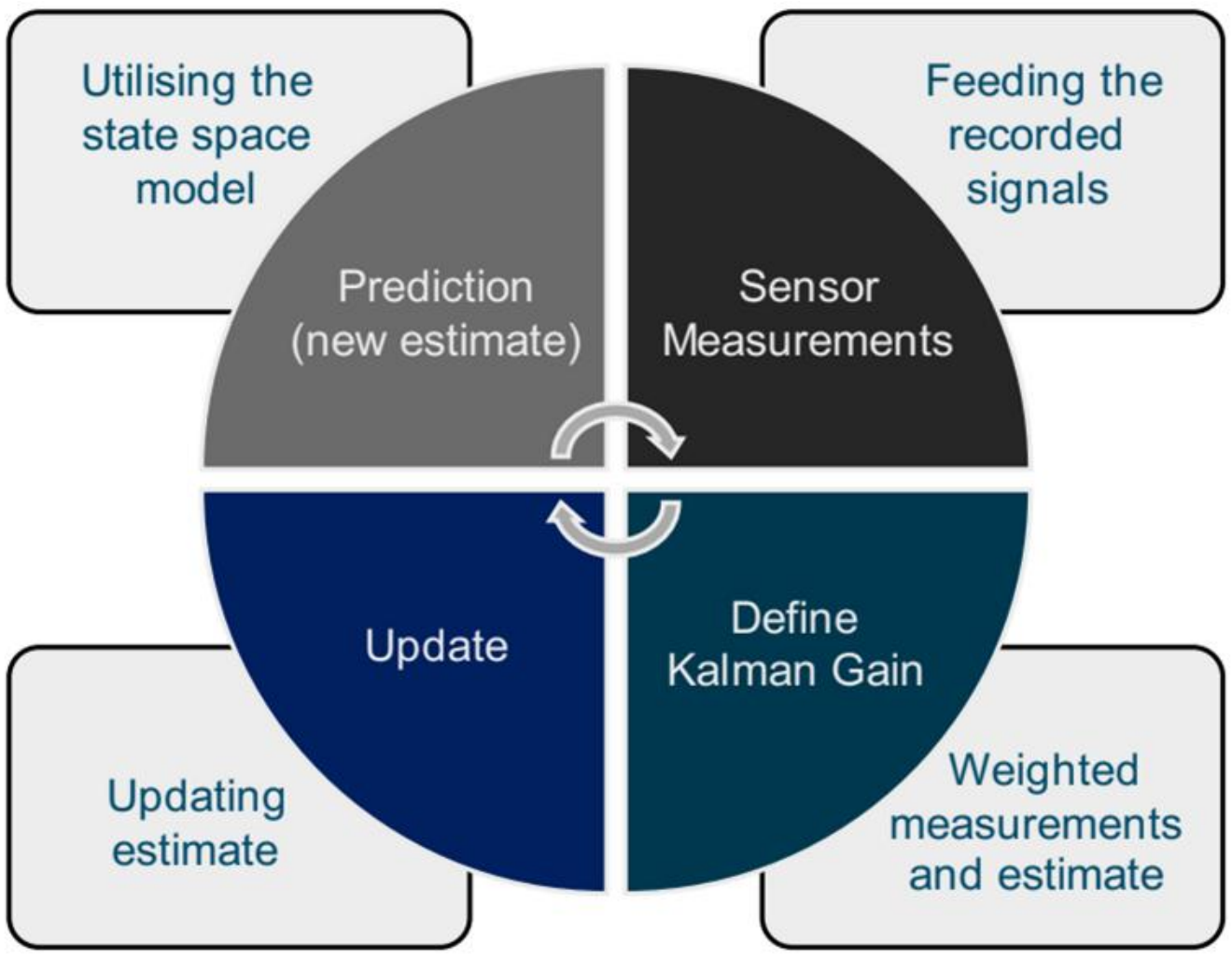

**Figure 7. The general framework of Kalman Filtering**

In GPR applications, the Kalman Filter (KF) has been used to estimate the background signal of radargrams and separate target signals from noise, as first proposed by Carevic (1999). This approach involves a stepwise analysis of the backscattered signal to estimate the internal condition of the system at specific times, referred to as the system state. The algorithm uses a parameter called "innovation," which represents the residual between actual and estimated signal values, to determine whether a target signal is present. This decision is made using the chi-squared ($\chi^2$) hypothesis test, a statistical method for comparing a model to observed data (McHugh, 2012). The target detection probability is based on the innovation values and the chi-squared test (Carevic, 1999; Luo and Fang, 2005; Ng et al., 2008; Zoubir et al., 2002). However, the algorithm has some limitations (Afrasiabi et al., 2025, 2023; Gurbuz, 2012):

- Lack of a robust algorithm to identify target width
- Absence of a definite method to determine the optimal threshold value for the chi-squared hypothesis
- Manual determination of input parameters, such as system and sensor noise covariances.

To either exploit the KF or address these limitations, several research studies in improving the detection and characterisation of buried infrastructure have been conducted which are presented in Table 88, specifically (Afrasiabi et al., 2025, 2023) which has focused on optimising KF parameters using evolutionary algorithms. Their algorithm, have shown possibility in automating the optimisation process, potentially improving the convergence and accuracy of KF performance in GPR applications.

**Table 8. Summary of works on GPR in correlation with Kalman filtering**

| | Article | Approaches | Purpose of utilising the KF |
|---|---|---|---|
| **1** | (Carevic, 1999) | ▪ Kalman Filter<br>▪ Chi-squared<br>▪ Background Adaption | ▪ Target detection and locating in GPR radargrams |
| **2** | (Van Kempen and Sahli, 2001b) | ▪ ARMA-Model<br>▪ Standard Kalman Filter | ▪ Utilised the parameters of the ARMA model to estimate clutters |
| **3** | (Ng et al., 2008) | ▪ Kalman Filter<br>▪ Chi-squared<br>▪ Background removal<br>▪ Matched Filter<br>▪ Deconvolution<br>▪ Wavelet Decomposition<br>▪ Trimmed Average Power | ▪ comparison of Carevic's KF performance in GPR application against other available methods for detection and locating buried targets |
| **4** | (Abeynayake et al., 2003) | ▪ Kalman filter<br>▪ Metal detector | ▪ landmine detection using metal detector arrays |
| **5** | (Luo and Fang, 2005) | ▪ Improved Kalman Filter<br>▪ Background removal<br>Improvement:<br>o Adaptive estimation<br>o Bayesian approach | ▪ Improvement of Carevic's KF for clutter reduction in GPR radargrams |
| **6** | (Torrione and Collins, 2006) | ▪ low- latency KF | ▪ Tracking of temporal position of the air/ground response in landmine detection |
| **7** | (Venkatasubramanian and Leung, 2006) | ▪ Interacting multiple model (IMM) filter<br>▪ Kalman Filter | ▪ Performance comparison of both methods |
| **8** | (Ng et al., 2008) | ▪ Kalman Filter<br>▪ Background removal<br>▪ Particle filter: | ▪ Comparison of particle filter, as a complex version of the KF, with Carevic's KF in landmine detection |
| | | o Reversible jump Markov chain Monte Carlo<br>o Sequential Monte Carlo | |
| **9** | (Smitha and Singh, 2016) | ▪ Kalman Filter<br>▪ Likelihood ratio test<br>▪ Singular Value Decomposition<br>▪ Principal Component Analysis<br>▪ Independent Component Analysis (ICA)<br>▪ Wavelet Transform (WT) | ▪ Applied Carevic's KF to reduce noise in GPR radargrams |
| **10** | (Dou et al., 2017) | ▪ Extended Kalman Filter (EKF) | ▪ Marching utility tracks in B-scans |
| **11** | (Xu et al., 2018) | ▪ Kalman filter<br>▪ Geometric correction<br>▪ Amplitude matching | ▪ Applied Carevic's KF to reduce noise in GPR radargrams |
| **12** | (Kaniewski and Kraszewski, 2020) | ▪ Kalman Filter<br>▪ Pendulum motion model | ▪ Position estimation of GPR antenna on the surface |
| **13** | (Zou et al., 2020) | ▪ Kalman Filter<br>▪ GPS | ▪ Tracking the position of the GPR antenna on the surface |
| **14** | (Afrasiabi et al., 2023) | ▪ Kalman Filter<br>▪ Genetic Algorithm<br>▪ Fourier Transform | ▪ Noise cluttering of GPR data and pipe detection using a semi-autonomous algorithm |
| **15** | (Afrasiabi et al., 2025) | ▪ Kalman Filter<br>▪ Genetic Algorithm<br>▪ Wavelet Transform | ▪ Noise cluttering of GPR data and pipe detection using an autonomous algorithm |

## 11 Discussion and Conclusion

This review has explored several non-intrusive methods used for the detection and characterisation of buried infrastructure, with a focus on geophysical techniques such as Electrical Resistivity Tomography (ERT), Infrared Thermography (IRT), and magnetometry, as well as the integration of multi-sensor and data fusion techniques. These methods play a vital role in improving the accuracy and reliability of subsurface surveys, yet each comes with specific advantages and limitations that must be addressed to enhance their applicability in real-world scenarios.

Electrical Resistivity Tomography (ERT) is a well-established technique that offers high resolution in shallow subsurface investigations, particularly for detecting buried infrastructure such as pipes and cables. However, ERT is sensitive to soil heterogeneity and the presence of moisture, which can complicate the interpretation of results. The integration of other geophysical methods, such as Ground Penetrating Radar (GPR), could significantly enhance the interpretation of ERT data, as both techniques offer complementary information on subsurface structures.

Infrared Thermography (IRT), on the other hand, offers a non-invasive, fast, and cost-effective means of detecting anomalies in the subsurface by exploiting thermal differences. Despite its advantages, IRT is constrained by the need for significant temperature contrast between the target and its surroundings, limiting its effectiveness under certain environmental conditions. Advances in thermographic technology, such as the development of long-wave infrared sensors, may improve the sensitivity of IRT and expand its utility for subsurface investigations.

Magnetometry offers the advantage of providing omnidirectional measurements, which allows for the detection of buried infrastructure even when the target is not directly beneath the sensor. However, distinguishing between ferrous objects such as pipes and cables can be challenging, and burial depth estimation remains a significant limitation. Research into enhanced segmentation techniques, coupled with other data sources, could potentially improve the accuracy of magnetometry for detecting complex and slender targets.

Data fusion and multi-sensor integration techniques are critical in overcoming the inherent limitations of individual sensing methods. The application of Kalman filters (KF) and other mathematical estimators, such as Particle Filters and Markov Chain Monte Carlo (MCMC) methods, holds great promise in improving data quality and reducing uncertainty in subsurface mapping. By integrating data from various sensors, these techniques allow for more robust decision-making and provide a more accurate representation of buried infrastructure. Nevertheless, the computational complexity of these methods, especially in real-time applications, remains a challenge that needs to be addressed.

While the discussed techniques have shown significant potential, their integration into practical applications still requires further validation under varied subsurface conditions. Future research should focus on optimising data fusion algorithms, enhancing sensor performance, and exploring new methodologies for integrating geophysical and remote sensing data.

In conclusion, the use of non-intrusive geophysical techniques for detecting and characterising buried infrastructure has proven to be a valuable tool in urban planning, environmental monitoring, and civil engineering projects. Methods such as ERT, IRT, and magnetometry, while effective on their own, each have inherent limitations that hinder their performance in certain environments or conditions. The integration of these techniques through data fusion approaches offers a promising solution, providing a more comprehensive and accurate view of the subsurface.

Mathematical estimators such as Kalman filters and their variations, including the Extended Kalman Filter (EKF) and Unscented Kalman Filter (UKF), have been shown to significantly improve data processing in GPR applications and other geophysical surveys. These estimators reduce noise, handle non-linearities, and enhance the accuracy of subsurface imaging, making them indispensable tools in modern geophysical investigations.

The key challenges moving forward include optimising sensor performance, refining data fusion techniques, and developing robust algorithms for real-time data processing. Future research should explore hybrid models that combine different geophysical methods, sensor technologies, and computational tools, such as deep learning algorithms, to further improve the precision and reliability of buried infrastructure detection.

In summary, while much progress has been made in the field of non-intrusive subsurface investigations, continuous innovation and interdisciplinary research are essential to advancing these technologies. The successful application of these methods could transform the way we monitor and maintain underground infrastructure, contributing to safer, more sustainable urban environments.

## Declarations

The authors have no conflicts of interest to declare that are relevant to the content of this article.

## Funding

Funding for this work was provided by the Engineering and Physical Sciences Research Council (EPSRC), award no. 2428310.

## Data and Code Availability Statement

No new datasets or custom-developed software were generated or analysed specifically for this review manuscript. All references cited provide relevant information regarding the datasets and methodologies discussed. Further details can be obtained directly from the cited literature.

## References

Abdelgwad, A.H., Nashat, A.A., 2017. Investigation of Utilizing L-Band Horn Antenna in Landmine Detection.

Abdelnour, A., Lazaro, A., Villarino, R., Kaddour, D., Tedjini, S., Girbau, D., 2018. Passive Harmonic RFID System for Buried Assets Localization. Sensors 18, 3635.

Abeynayake, C.G., Chant, I.J., Nash, G., 2003. Kalman detection of landmines in metal detector array data. Detect. Remediat. Technol. Mines Minelike Targets VIII 5089, 1243.

Abraham, D.A., 2017. Signal Processing. In: Applied Underwater Acoustics: Leif Bjørnø. Elsevier, pp. 743–807.

Adamopoulos, E., Colombero, C., Comina, C., Rinaudo, F., Volinia, M., Girotto, M., Ardissono, L., 2021. Integrating multiband photogrammetry, scanning, and GPR for built heritage surveys: The façades of Castello del Valentino. ISPRS Ann. Photogramm. Remote Sens. Spat. Inf. Sci. 8, 1–8.

Afrasiabi, A., 2023. Novel AI-assisted computational solutions for GPR data interpretation and electromagnetic data fusion to detect buried utilities By A thesis submitted to The University of Birmingham School of Engineering. Birmingham.

Afrasiabi, A., Faramarzi, A., Chapman, D., Keshavarzi, A., 2025. Optimising Ground

Penetrating Radar data interpretation: A hybrid approach with AI-assisted Kalman Filter and Wavelet Transform for detecting and locating buried utilities. J. Appl. Geophys. 232, 105567.

Afrasiabi, A., Faramarzi, A., Chapman, D., Keshavarzi, A., Stringfellow, M., 2023. Toward the optimisation of the Kalman Filter approach in ground penetrating radar application for detection and locating buried utilities. J. Appl. Geophys. 219, 105220.

Ahn, S.J., Rauh, W., Warnecke, H.-J., 2001. Least-squares orthogonal distances fitting of circle, sphere, ellipse, hyperbola, and parabola. Pattern Recognit. 34, 2283–2303.

Akagi, H., Tamura, S., 2006. A passive EMI filter for eliminating both bearing current and ground leakage current from an inverter-driven motor. IEEE Trans. Power Electron. 21, 1459–1468.

Al-Habaibeh, A., Anderson, G., Damji, K., Jones, G., Lam, K., 2010. Using infrared thermography for monitoring thermal efficiency of buildings - case studies from Nottingham Trent University.

Al-Nuaimy, W., 1999. Automatic feature detection and interpretation in ground-penetrating radar data. The University of Liverpool.

Al-Nuaimy, W., Huang, Y., Nakhkash, M., Fang, M.T.C., Nguyen, V.T., Eriksen, A., 2000. Automatic detection of buried utilities and solid objects with GPR using neural networks and pattern recognition, Journal of Applied Geophysics.

Al-Nuaimy, W., Shihab, S., Eriksen, A., 2004. Data fusion for accurate characterisation of buried cylindrical objects using GPR. In: Proceedings of the Tenth International Conference on Ground Penetrating Radar. pp. 359–362.

Ali, H., Ahmad Zaidi, A.F., Wan Ahmad, W.K., Zanar Azalan, M.S., Tengku Amran, T.S., Ahmad, M.R., Elshaikh, M., 2021. A cascade hyperbolic recognition of buried objects using hybrid feature extraction in ground penetrating radar images. In: Journal of Physics: Conference Series. IOP Publishing, p. 012018.

Alipio, R., Duarte, N., Karami, H., Rubinstein, M., Rachidi, F., 2023. Assessment of the feasibility of applying the electromagnetic time reversal theory to locate defects in grounding electrodes. Energies 16, 5104.

Allred, B.J., Daniels, J.J., Fausey, N.R., Chen, C., Peters, L., Youn, H., 2013. Important considerations for locating buried agricultural drainage pipe using ground penetrating radar. Appl. Eng. Agric. 21, 71–87.

Amaral, L.C.M., Roshan, A., Bayat, A., 2022. Review of Machine Learning Algorithms for Automatic Detection of Underground Objects in GPR Images. J. Pipeline Syst. Eng. Pract. 13, 04021082.

Aslam, H., Kaur, M., Sasi, S., Mortula, M.M., Yehia, S., Ali, T., 2018. Detection of leaks in water distribution system using non-destructive techniques. In: IOP Conference Series: Earth and Environmental Science. IOP Publishing, p. 12004.

Atef, A., Zayed, T., Hawari, A., Khader, M., Moselhi, O., 2016. Multi-tier method using infrared photography and GPR to detect and locate water leaks. Autom. Constr. 61, 162–170.

Baek, J., Yoon, J.S., Lee, C.M., Choi, Y., 2018. A Case Study on Detection of Subsurface

Cavities of Urban Roads Using Ground-coupled GPR. In: 2018 17th International Conference on Ground Penetrating Radar, GPR 2018. Institute of Electrical and Electronics Engineers Inc.

Bai, H., Sinfield, J. V., 2020. Improved background and clutter reduction for pipe detection under pavement using Ground Penetrating Radar (GPR). J. Appl. Geophys. 172, 103918.

Bar-Shalom, Y., Li, X.-R., Kirubarajan, T., 2001. Estimation with Applications to Tracking and Navigation, Estimation with Applications to Tracking and Navigation.

Bárdossy, A., Singh, S.K., 2008. Robust estimation of hydrological model parameters. Hydrol. earth Syst. Sci. 12, 1273–1283.

Beamish, D., 2011. Low induction number, ground conductivity meters: A correction procedure in the absence of magnetic effects. J. Appl. Geophys. 75, 244–253.

Belotti, V., Dell'Acqua, F., Gamba, P., 2003. Denoising and Hyperbola Recognition in GPR data. Image Signal Process. Remote Sens. VII 4541, 70–80.

Benedetto, A., Benedetto, F., 2014. Application Field-Specific Synthesizing of Sensing Technology: Civil Engineering Application of Ground-Penetrating Radar Sensing Technology. In: Comprehensive Materials Processing. Elsevier, pp. 393–425.

Benedetto, A., Tosti, F., Bianchini Ciampoli, L., D'Amico, F., 2017. An overview of ground-penetrating radar signal processing techniques for road inspections-2. Signal Processing 132, 201–209.

Bentley, L.R., Trenholm, N.M., 2002. The Accuracy of Water Table Elevation Estimates Determined from Ground Penetrating Radar Data. J. Environ. Eng. Geophys. 7, 37–53.

Bi, W., Zhao, Y., An, C., Hu, S., 2018. Clutter elimination and random-noise denoising of GPR signals using an SVD method based on the Hankel matrix in the local frequency domain. Sensors (Switzerland) 18, 3422.

Bianchini Ciampoli, L., Gagliardi, V., Clementini, C., Latini, D., Del Frate, F., Benedetto, A., 2020. Transport Infrastructure Monitoring by InSAR and GPR Data Fusion. Surv. Geophys.

Bigman, D.P., 2012. The use of electromagnetic induction in locating graves and mapping cemeteries: An Example from native North America. Archaeol. Prospect. 19, 31–39.

Billings, S.D., 2004. Discrimination and classification of buried unexploded ordnance using magnetometry. IEEE Trans. Geosci. Remote Sens. 42, 1241–1251.

Bimpas, M., Amditis, A., Uzunoglu, N., 2010. Detection of water leaks in supply pipes using continuous wave sensor operating at 2.45 GHz. J. Appl. Geophys. 70, 226–236.

Bimpas, M., Amditis, A., Uzunoglu, N.K., 2011. Design and implementation of an integrated high resolution imaging ground penetrating radar for water pipeline rehabilitation. Water Resour. Manag. 25, 1239–1250.

bin Haron, A.H., Ahmad, M.R. Bin, Masenwat, N.A., Sani, S., Mustapha, I., Sayuti, S., Razak, N.B.A., Adnan, M.A.K. Bin, bin Mohd Yusoff, M.A., bin Mohsin, M.A.D., 2024. Subsurface utility detection using ground penetrating radar and electromagnetic locator– a comparative study. In: IOP Conference Series: Materials Science and Engineering. IOP Publishing, p. 12013.

Binder, D., Bru¨cklbru¨bru¨ckl, E., Roch, K.H., Behm, M., Scho¨ner, W., Scho¨ner, S., Hynek, B., 2020. Determination of total ice volume and ice-thickness distribution of two glaciers in the Hohe Tauern region, Eastern Alps, from GPR data.

Birkenfeld, S., 2010. Automatic detection of reflexion hyperbolas in gpr data with neural networks. World Autom. Congr. (WAC), 2010 1–6.

Borgioli, G., Capineri, L., Falorni, P., Matucci, S., Windsor, C.G., 2008. The detection of buried pipes from time-of-flight radar data. IEEE Trans. Geosci. Remote Sens. 46, 2254–2266.

Bray, D.E., Stanley, R.K., 2018. Nondestructive evaluation: a tool in design, manufacturing and service. CRC press.

Bricker, S., Frith, N., 2017. Mapping Underground 11–12.

Brunton, S.L., Kutz, J.N., 2017. Data Driven Science & Engineering - Machine Learning, Dynamical Systems, and Control.

Bryakin, I. V, Bochkarev, I. V, 2019. Cable Avoidance Tool.

Bu, Z., Gao, G., Li, H.J., Cao, J., 2017. CAMAS: A cluster-aware multiagent system for attributed graph clustering. Inf. Fusion 37, 10–21.

Busby, J.P., Merritt, J.W., 1999. Quaternary deformation mapping with ground penetrating radar. J. Appl. Geophys. 41, 75–91.

Byrd, R.H., Lu, P., Nocedal, J., Zhu, C., 1995. A limited memory algorithm for bound constrained optimization. SIAM J. Sci. Comput. 16, 1190–1208.

Cagniard, L., 1953. Basic theory of the magneto-telluric method of geophysical prospecting. Geophysics 18, 605–635.

Cao, J., Wu, Z., Wu, J., Liu, W., 2013. Towards information-theoretic K-means clustering for image indexing. Signal Processing 93, 2026–2037.

Capineri, L., Grande, P., Temple, J.A.G., 1998. Advanced image-processing technique for real-time interpretation of ground-penetrating radar images. Int. J. Imaging Syst. Technol. 9, 51–59.

Capozzoli, L., Rizzo, E., 2017. Combined NDT techniques in civil engineering applications: Laboratory and real test. Constr. Build. Mater. 154, 1139–1150.

Carevic, D., 1999. Kalman filter-based approach to target detection and target-background separtion in ground-penetrating radar data. In: Detection and Remediation Technologies for Mines and Minelike Targets IV. p. 1284.

Carreño-Alvarado, E.P., Ayala-Cabrera, D., Pérez-García, R., Izquierdo, J., 2014. Identification of buried pipes using thermal images and data mining. Procedia Eng. 89, 1445–1451.

Carrick Utsi, E., 2017a. Examples of Practical Problems. Gr. Penetrating Radar 135–151.

Carrick Utsi, E., 2017b. Multichannel and Single Channel Systems. Gr. Penetrating Radar 153–160.

Cassidy, N., 2009. Ground penetrating radar data processing, modelling and analysis, First Edit. ed, Ground Penetrating Radar. Elsevier.

Chalikakis, K., Plagnes, V., Guerin, R., Valois, R., Bosch, F.P., 2011. Contribution of geophysical methods to karst-system exploration: An overview. Hydrogeol. J. 19, 1169–1180.

Che Ku Melor, C.K.N.A.H., Joret, A., Razali, M., Ponniran, A., Sulong, M.S., Omar, R., 2021. Frequency based signal processing technique for pulse modulation ground penetrating radar system. Int. J. Electr. Comput. Eng. 11, 4104–4112.

Chen, D.H., Wimsatt, A., 2010. Inspection and condition assessment using ground penetrating radar. J. Geotech. Geoenvironmental Eng. 136, 207–214.

Chen, H.-W., Huang, T.-M., 1998. Finite-difference time-domain simulation of GPR data. J. Appl. Geophys. 40, 139–163.

Chen, H., 2010. Probabilistic conic mixture model and its applications to mining spatial ground penetrating radar data. SIAM Conf. Data Min. (.

Christensen, N.B., Tølbøll, R.J., 2009. A lateral model parameter correlation procedure for one-dimensional inverse modelling. Geophys. Prospect. 57, 919–929.

Churchill, K.M., Link, C., Youmans, C.C., 2012. A Comparison of the Finite-Element Method and Analytical Method for Modeling Unexploded Ordnance Using Magnetometry. IEEE Trans. Geosci. Remote Sens. 50, 2720–2732.

Ciampoli, L.B., Tosti, F., Economou, N., Benedetto, F., 2019. Signal Processing of GPR Data for Road Surveys. Geosciences 9, 96.

Collins, L.M., Huettel, L.G., Simpson, W.A., Tantum, S.L., 2002. Sensor fusion of EMI and GPR data for improved land mine detection. In: Detection and Remediation Technologies for Mines and Minelike Targets VII. pp. 872–879.

Conyers, L.B., Lucius, J.E., 1996. Velocity Analysis in Archaeological Ground-Penetrating Radar Studies. Archaeol. Prospect. 3, 25–38.

Daniels, D.J., 2004. Ground Penetrating Radar-2nd edition, 2nd ed, IEE Radar Series. The Institution of Electrical Engineers, London, United Kingdom, London.

Davis, C., Knowles, M., Rajic, N., Swanton, G., 2016. Evaluation of a distributed fibre optic strain sensing system for full-scale fatigue testing. Procedia Struct. Integr. 2, 3784–3791.

Dawalibi, F., 1986. Electromagnetic fields generated by overhead and buried short conductors part 2 - ground networks. IEEE Trans. Power Deliv. 1, 112–119.

De Coster, A., Pérez Medina, J.L., Nottebaere, M., Alkhalifeh, K., Neyt, X., Vanderdonckt, J., Lambot, S., 2019. Towards an improvement of GPR-based detection of pipes and leaks in water distribution networks. J. Appl. Geophys. 162, 138–151.

De Coster, A., Van der Wielen, A., Grégoire, C., Lambot, S., 2018. Evaluation of pavement layer thicknesses using GPR: A comparison between full-wave inversion and the straight-ray method. Constr. Build. Mater. 168, 91–104.

Denisov, A.Y., Denisova, O. V., Sapunov, V.A., Khomutov, S.Y., 2006. Measurement quality estimation of proton-precession magnetometers. Earth, Planets Sp. 58, 707–710.

Dewantara, D., Parnadi, W.W., 2022. Automatic Hyperbola Detection and Apex Extraction Using Convolutional Neural Network on GPR Data. In: Journal of Physics: Conference Series. IOP Publishing, p. 012027.

Dey, A., Morrison, H.F., 1979. Resistivity Modeling for Arbitrarily Shaped Three-Dimensional Structures. Geophysics 44, 753–780.

Dojack, L., 2012. Ground Penetrating Radar Theory, Data Collection, Processing, and Interpretation: A Guide for Archaeologists. Proj. Report, Univ. Br. Columbia.

Doolittle, J.A., Brevik, E.C., 2014. The use of electromagnetic induction techniques in soils studies. Geoderma.

Dossi, M., Forte, E., Pipan, M., 2015. Automated reflection picking and polarity assessment through attribute analysis: Theory and application to synthetic and real ground-penetrating radar data. Geophysics 80, H23–H35.

Dou, Q., Wei, L., Magee, D.R., Cohn, A.G., 2017. Real-Time Hyperbola Recognition and Fitting in GPR Data. IEEE Trans. Geosci. Remote Sens. 55, 51–62.

Downs, C., Jazayeri, S., 2021. Resolution enhancement of deconvolved ground penetrating radar images using singular value decomposition. J. Appl. Geophys. 193, 104401.

Duan, Q., Sorooshian, S., Gupta, V.K., 1994. Optimal use of the SCE-UA global optimization method for calibrating watershed models. J. Hydrol. 158, 265–284.

Duda, R.O., Hart, P.E., 2002. Use of the Hough transformation to detect lines and curves in pictures. Commun. ACM 15, 11–15.

Dutta, R., Cohn, A.G., Muggleton, J.M., 2013. 3D mapping of buried underworld infrastructure using dynamic Bayesian network based multi-sensory image data fusion. J. Appl. Geophys. 92, 8–19.

Economou, N., Benedetto, F., Bano, M., Tzanis, A., Nyquist, J., Sandmeier, K.-J., Cassidy, N., 2017. Advanced Ground Penetrating Radar Signal Processing Techniques. Signal Processing 132, 197–200.

El Afi, M., Belattar, S., 2014. Application of the Infrared Thermography in Detection of Pipes in Concrete Structures.

Eppstein, M.J., Dougherty, D.E., 1998. and moisture monitoring from cross-well ground-penetrating radar test site Vermont phase liquid ( DNAPL ) plume month ; Water Resour. Res. 34, 1889–1900.

Ertöz, L., Steinbach, M., Kumar, V., 2013. Finding Clusters of Different Sizes, Shapes, and Densities in Noisy, High Dimensional Data 47–58.

Everett, M.E., 2013. Near-Surface Applied Geophysics. Cambridge University Press.

Fahmy, M., Moselhi, O., 2017. Detecting and Locating Leaks in Underground Water Mains Using Thermography. Proc. 2009 Int. Symp. Autom. Robot. Constr. (ISARC 2009) 61–67.

Farzamian, M., Paz, M.C., Paz, A.M., Castanheira, N.L., Gonçalves, M.C., Monteiro Santos, F.A., Triantafilis, J., 2019. Mapping soil salinity using electromagnetic conductivity imaging—A comparison of regional and location-specific calibrations. L. Degrad. Dev. 30, 1393–1406.

Faundez-Zanuy, M., 2009. Data fusion at different levels. In: Lecture Notes in Computer Science (Including Subseries Lecture Notes in Artificial Intelligence and Lecture Notes in Bioinformatics). Springer, Berlin, Heidelberg, pp. 94–103.

Feng, S., Liu, D., Cheng, X., Fang, H., Li, C., 2017. A new segmentation strategy for processing magnetic anomaly detection data of shallow depth ferromagnetic pipeline. J. Appl. Geophys. 139, 65–72.

Fischer, G., Fraedrich, V., Wendel, C., 2002. Method and device for locating a metal line. US 6,411,073 B1.

Fletcher, R., Reeves, C.M., 1964. Function minimization by conjugate gradients. Comput. J. 7, 149–154.

Galetti, E., Curtis, A., 2018. Transdimensional electrical resistivity tomography. J. Geophys. Res. Solid Earth 123, 6347–6377.

Gallagher, K., Charvin, K., Nielsen, S., Sambridge, M., Stephenson, J., 2009. Markov chain Monte Carlo (MCMC) sampling methods to determine optimal models, model resolution and model choice for Earth Science problems. Mar. Pet. Geol. 26, 525–535.

Gamba, P., Belotti, V., 2003. Two fast buried pipe detection schemes in Ground Penetrating Radar images. Int. J. Remote Sens. 24, 2467–2484.

Gamba, P., Lossani, S., 2000. Neural detection of pipe signatures in ground penetrating radar images. IEEE Trans. Geosci. Remote Sens. 38, 790–797.

Ganiyu, S.A., Oladunjoye, M.A., Onakoya, O.I., Olutoki, J.O., Badmus, B.S., 2020. Combined electrical resistivity imaging and ground penetrating radar study for detection of buried utilities in Federal University of Agriculture, Abeokuta, Nigeria. Environ. Earth Sci. 79, 1–20.

García-Escudero, L.A., Mayo-Iscar, A., Sánchez-Gutiérrez, C.I., 2017. Fitting parabolas in noisy images. Comput. Stat. Data Anal. 112, 80–87.

Giannakis, I., Giannopoulos, A., Warren, C., 2019. A Machine Learning-Based Fast-Forward Solver for Ground Penetrating Radar with Application to Full-Waveform Inversion. IEEE Trans. Geosci. Remote Sens. 57, 4417–4426.

Giannakis, I., Tosti, F., Lantini, L., Alani, A.M., 2020. Diagnosing Emerging Infectious Diseases of Trees Using Ground Penetrating Radar. IEEE Trans. Geosci. Remote Sens. 58, 1146–1155.

Goddard, K.F., Wang, P., Lewin, P.L., Swingler, S.G., 2012. Detection and location of underground cables using magnetic field measurements. Meas. Sci. Technol. 23, 055002.

Gonzalez-DIaz, M., Garcia-Fernandez, M., Alvarez-Lopez, Y., Las-Heras, F., 2020. Improvement of GPR SAR-Based Techniques for Accurate Detection and Imaging of Buried Objects. IEEE Trans. Instrum. Meas. 69, 3126–3138.

Goodman, D., Piro, S., 2013a. GPR remote sensing in archaeology. GPR Remote Sens. Archaeol. 1–233.

Goodman, D., Piro, S., 2013b. GPR remote sensing in archaeology -Chapter2. In: GPR Remote Sensing in Archaeology. pp. 1–233.

Grandjean, G., Leparoux, D., 2016. The potential of seismic methods for detecting cavities and buried objects: experimentation at a test site.

Greaves, R.J., Lesmes, D.P., Lee, J.M., Toksöz, M.N., 1996. Velocity variations and water content estimated from multi-offset, ground-penetrating radar. Geophysics 61, 683–695.

Greggio, N., Giambastiani, B.M.S., Balugani, E., Amaini, C., Antonellini, M., 2018. High-resolution electrical resistivity tomography (ERT) to characterize the spatial extension of freshwater lenses in a salinized coastal aquifer. Water (Switzerland) 10, 1067.

Gros, 1997a. NDT data fusion. Arnold, London.

Gros, 1997b. Data Fusion – A Review. Butterworth-Heinemann, Oxford, pp. 5–42.

Guillemoteau, J., Christensen, N.B., Jacobsen, B.H., Tronicke, J., 2017. Fast 3D multichannel deconvolution of electromagnetic induction loop-loop apparent conductivity data sets acquired at low induction numbers. Geophysics 82, E357–E369.

Guo, Z.-Y., Liu, D.-J., Pan, Q., Zhang, Y.-Y., 2015. Forward modeling of total magnetic anomaly over a pseudo-2D underground ferromagnetic pipeline. J. Appl. Geophys. 113, 14–30.

Gurbuz, A.C., 2012. Determination of background distribution for ground-penetrating radar data. IEEE Geosci. Remote Sens. Lett. 9, 544–548.

Hafsi, M., Bolon, P., Dapoigny, R., 2017. Detection and localization of underground networks by fusion of electromagnetic signal and GPR images. In: Nagahara, H., Umeda, K., Yamashita, A. (Eds.), Thirteenth International Conference on Quality Control by Artificial Vision 2017. SPIE, p. 1033803.

Hajipour, S., Azadi Namin, F., Sarraf Shirazi, R., 2023. A novel method for GPR imaging based on neural networks and dictionary learning. Waves in Random and Complex Media 33, 393–413.

Hao, T., Rogers, C.D.F., Metje, N., Chapman, D.N., Muggleton, J.M., Foo, K.Y., Wang, P., Pennock, S.R., Atkins, P., Swingler, S.G., Parker, J., Costello, S.B., Burrow, M.P.N., Anspach, J.H., Armitage, R.J., Cohn, A.G., Goddard, K., Lewin, P.L., Orlando, G., Redfern, M.A., Royal, A.C.D., Saul, A.J., 2012. Condition assessment of the buried utility service infrastructure. Tunn. Undergr. Sp. Technol. 28, 331–344.

Harkat, H., Elfakir, Y., Bennani, S.D., Khaissidi, G., Mrabti, M., 2016. Ground penetrating radar hyperbola detection using Scale-Invariant Feature Transform. In: Proceedings of 2016 International Conference on Electrical and Information Technologies, ICEIT 2016. IEEE, pp. 392–397.

Harkat, H., Ruano, A., Ruano, M.G., Bennani, S.D., 2018. Classifier Design by a Multi-Objective Genetic Algorithm Approach for GPR Automatic Target Detection. IFAC-PapersOnLine 51, 187–192.

Harkat, H., Ruano, A.E., Ruano, M.G., Bennani, S.D., 2019. GPR target detection using a neural network classifier designed by a multi-objective genetic algorithm. Appl. Soft Comput. J. 79, 310–325.

Hart, P., Duda, R.O., 1972. Use of the hough trasformtion to detect lines and curves in pictures. Commun. ACM 15, 11–15.

Herter, C., 2019. The electromagnetic spectrum: A critical natural resource. In: Transboundary Resources Law. pp. 89–101.

Ho, K.C., Gader, P.D., Wilson, J.N., 2004. Improving landmine detection using frequency domain features from ground penetrating radar. In: International Geoscience and Remote Sensing Symposium (IGARSS). pp. 1617–1620.

Holt, P., 2014. Marine Magnetometer Processing. 3H Consulting, Plymouth:

Holtham, E., Marchant, D., McMillan, M., Haber, E., 2017. Advanced Modelling of Electromagnetic Data. Proc. Explor. 17 Sixth Decenn. Int. Conf. Miner. Explor. 209–219.

Huang, X., Wang, W., Lv, Z., Meng, Y., Yuan, C., Feng, J., 2019. Precise Positioning of the Underground Power Cable by Magnetic Field Detectio. In: Annual Report - Conference on Electrical Insulation and Dielectric Phenomena, CEIDP. Institute of Electrical and Electronics Engineers Inc., pp. 702–705.

Huber, E., Hans, G., 2018. RGPR - An open-source package to process and visualize GPR data. In: 2018 17th International Conference on Ground Penetrating Radar, GPR 2018. Institute of Electrical and Electronics Engineers Inc.

Huray, P.G., 2010. Maxwell ' s Equations, IEEE press. Wiley.

Idi, B.Y., Kamarudin, M.N., 2012a. Interpretation of ground penetrating radar image using digital wavelet transform. Asian J. Appl. Sci. 5, 174–182.

Idi, B.Y., Kamarudin, M.N., 2012b. Interpretation of ground penetrating radar image using digital wavelet transform. Asian J. Appl. Sci. 5, 174–182.

Iswandy, A., Serma, A., Setan, H., 2009. Ground penetrating radar (GPR) for subsurface mapping: preliminary result. Geoinf. Sci. J. 9, 45–62.

Jaganathan, A.P., Allouche, E., Simicevic, N., 2010. Numerical modeling and experimental evaluation of a time domain UWB technique for soil void detection. Tunn. Undergr. Sp. Technol. 25, 652–659.

Jeong, H.S., Arboleda, C.A., Abraham, D.M., Halpin, D.W., Bernold, L.E., 2003. Imaging and Locating Buried Utilities. Technol. Transf. Proj. Implement. Inf. 7, 65,66.

Jiles, D.C., 1990. Review of magnetic methods for nondestructive evaluation (Part 2). NDT Int. 23, 83–92.

Jol, H.M., 2009. Ground Penetrating Radar Theory and Application, Ground Penetrating Radar. Elsevier.

Junoh, M.S.A.M., Sulaiman, S.A.H., Natnan, S.R., Purwanto, H., 2022. Estimation diameter of buried pipe using principle of ground penetrating radar and electromagnetic locator. Int. J. Geoinformatics 18, 61–72.

Kadioglu, S., 2018. GPR Data Visualization Enhancement by Deconvolution and Some Image Attributes. In: 2018 41st International Conference on Telecommunications and Signal Processing, TSP 2018. Institute of Electrical and Electronics Engineers Inc.

Kalika, D., Morton, K.D., Collins, L.M., Torrione, P.A., 2014. Hyperbolic and PLSDA filter algorithms to detect buried threats in GPR data. In: Bishop, S.S., Isaacs, J.C. (Eds.), Detection and Sensing of Mines, Explosive Objects, and Obscured Targets XIX. International Society for Optics and Photonics, p. 90720U.

Kaniewski, P., Kraszewski, T., 2020. Novel algorithm for position estimation of handheld ground-penetrating radar antenna. Proc. Int. Radar Symp. 2020-Octob, 100–102.

Kanungo, T., Mount, D.M., Netanyahu, N.S., Piatko, C.D., Silverman, R., Wu, A.Y., 2002. An efficient k-means clustering algorithms: Analysis and implementation. IEEE Trans. Pattern Anal. Mach. Intell. 24, 881–892.

Karamitrou, A., Bogiatzis, P., Tsokas, G.N., 2020. Fusion of geophysical images in the study of archaeological sites. Archaeol. Prospect. 27, 119–133.

Kemna, A., Kulessa, B., Vereecken, H., 2002. Imaging and characterisation of subsurface solute transport using electrical resistivity tomography (ERT) and equivalent transport models. J. Hydrol. 267, 125–146.

Khamzin, A.K., Varnavina, A. V., Torgashov, E. V., Anderson, N.L., Sneed, L.H., 2017. Utilization of air-launched ground penetrating radar (GPR) for pavement condition assessment. Constr. Build. Mater. 141, 130–139.

Knox, M., Rundel, C., Collins, L., 2017. Sensor fusion for buried explosive threat detection for handheld data. In: Detection and Sensing of Mines, Explosive Objects, and Obscured Targets XXII. p. 101820D.

Kofman, L., Ronen, A., Frydman, S., 2006. Detection of model voids by identifying reverberation phenomena in GPR records. J. Appl. Geophys. 59, 284–299.

Kovalenko, V., Masalov, S.A., 2000. 2D matrix filtering of ground penetrating radar data. In: International Conference on Mathematical Methods in Electromagnetic Theory, MMET. IEEE Computer Society, pp. 236–238.

Ku, C.C., 1976. Numerical Inverse Magnetotelluric Problems. Geophysics 41, 276–286.

Kumar, D., 2012. Efficacy of electrical resistivity tomography technique in mapping shallow subsurface anomaly. J. Geol. Soc. India 80, 304–307.

Kumlu, D., Erer, I., Kaplan, N.H., 2020. Low complexity clutter removal in GPR images via lattice filters. Digit. Signal Process. A Rev. J. 101, 102724.

Lai, W.W.L., Chang, R.K.W., Sham, J.F.C., 2018. A blind test of nondestructive underground void detection by ground penetrating radar (GPR). J. Appl. Geophys. 149, 10–17.

Lai, W.W.L., Chang, R.K.W., Sham, J.F.C., Pang, K., 2016. Perturbation mapping of water leak in buried water pipes via laboratory validation experiments with high-frequency ground penetrating radar (GPR).

Lambot, S., Moghadas, D., André, F., Slob, E.C., Vereecken, H., 2009. A unified full-waveform method for modeling ground penetrating radar and electromagnetic induction data for non-destructive characterization of soil and materials. In: Proceedings of the 2009 International Conference on Electromagnetics in Advanced Applications, ICEAA '09. pp. 1058–1061.

Lei, W., Hou, F., Xi, J., Tan, Q., Xu, M., Jiang, X., Liu, G., Gu, Q., 2019. Automatic hyperbola detection and fitting in GPR B-scan image. Autom. Constr. 106, 102839.

Lesparre, N., Robert, T., Nguyen, F., Boyle, A., Hermans, T., 2019. 4D electrical resistivity tomography (ERT) for aquifer thermal energy storage monitoring. Geothermics 77, 368–382.

Li, H., Chou, C., Fan, L., Li, B., Wang, D., Song, D., 2019. Toward Automatic Subsurface Pipeline Mapping by Fusing a Ground-Penetrating Radar and a Camera. IEEE Trans. Autom. Sci. Eng. 17, 722–734.

Li, S., Cai, H., Kamat, V.R., 2015. Uncertainty-aware geospatial system for mapping and visualizing underground utilities. Autom. Constr. 53, 105–119.

Li, Y., 2015. Detection and Identification of Subsurface Corrosion in Ferromagnetic Double-casing Pipes via Pulsed Remote Field Eddy Current Technique. In: Studies in Applied Electromagnetics and Mechanics. IOS Press, pp. 228–235.

Li, Y., Zhao, Z., Luo, Y., Qiu, Z., 2020. Real-time pattern-recognition of GPR images with YOLO V3 implemented by tensorflow. Sensors (Switzerland) 20, 1–17.

Liggins, M.E., Hall, D.L., Llinas, J., 2009. Handbook of Multisensor Data Fusion: Theory and Practice, CRC Press.

Ligthart, L., Yarovoy, A., 2003. STW project "Advanced relocatable multi-sensor system for buried landmine detection." In: Proceedings of the 2nd International Workshop on Advanced Ground Penetrating Radar.

Lim, M.K., Cao, H., 2013. Combining multiple NDT methods to improve testing effectiveness. Constr. Build. Mater. 38, 1310–1315.

Liu, Z., Gu, X., Chen, J., Wang, D., Chen, Y., Wang, L., 2023. Automatic recognition of pavement cracks from combined GPR B-scan and C-scan images using multiscale feature fusion deep neural networks. Autom. Constr. 146, 104698.

Liu, Z., Kleiner, Y., 2012. State-of-the-art review of technologies for pipe structural health monitoring. IEEE Sens. J.

Lu, G., Zhao, W., Forte, E., Tian, G., Li, Y., Pipan, M., 2020. Multi-frequency and multi-attribute GPR data fusion based on 2-D wavelet transform. Meas. J. Int. Meas. Confed. 166, 108243.

Luo, T., Zhu, S., Yikeremu, Y., Zhu, J., Genderen, J. van, 2023. Ground penetrating radar applied to subsurface culverts. Geo-Spatial Inf. Sci.

Luo, T.X.H., Lai, W.W.L., 2020. GPR pattern recognition of shallow subsurface air voids. Tunn. Undergr. Sp. Technol. 99, 103355.

Luo, T.X.H., Lai, W.W.L., Chang, R.K.W., Goodman, D., 2019. An empirical study of GPR imaging criteria. J. Appl. Geophys. 165, 37–48.

Luo, Y., Fang, G.Y., 2005. GPR clutter reduction and buried target detection by improved Kalman filter technique. In: 2005 International Conference on Machine Learning and Cybernetics, ICMLC 2005. IEEE, pp. 5432–5436.

Maas, C., Schmalzl, J., 2013. Using pattern recognition to automatically localize reflection hyperbolas in data from ground penetrating radar. Comput. Geosci. 58, 116–125.

Madden, T.R., Swift, C.M., 1965. Magnetotelluric Studies of the Electrical Conductivity Structure of the Crust and Upper Mantle. Earth's Crust Up. Mantle 9, 317–345.

Mahmoudzadeh Ardekani, M.R., 2013. Off- and on-ground GPR techniques for field-scale soil moisture mapping. Geoderma 200–201, 55–66.

Majewski, R.S., Valenta, J., Tábořík, P., Weger, J., Kučera, A., Patočka, Z., Čermák, J., 2022. Geophysical imaging of tree root absorption and conduction zones under field conditions: a comparison of common geoelectrical methods. Plant Soil 481, 447–473.

Makana, L., Metje, N., Jefferson, I.F., David, C., Rogers, F., 2016. What Do Utility Strikes Really Cost ? A Report by the University of Birmingham – School of Civil Engineering.

Makar, J., Chagnon, N., 1999. Inspecting systems for leaks, pits, and corrosion. J. / Am. Water Work. Assoc. 91, 36–46.

Mallat, S., 2009. A theory for multiresolution signal decomposition: The wavelet representation. In: Fundamental Papers in Wavelet Theory. pp. 494–513.

Manataki, M., Vafidis, A., Sarris, A., 2021. Gpr data interpretation approaches in archaeological prospection. Appl. Sci. 11, 7531.

Massaro, A., Savino, N., Selicato, S., Panarese, A., Galiano, A., Dipierro, G., 2021. Thermal IR and GPR UAV and vehicle embedded sensor non-invasive systems for road and bridge inspections. In: 2021 IEEE International Workshop on Metrology for Industry 4.0 and IoT, MetroInd 4.0 and IoT 2021 - Proceedings. Institute of Electrical and Electronics Engineers Inc., pp. 248–253.

Mazor, E., Averbuch, A., Bar-Shalom, Y., Dayan, J., 1998. Interacting multiple model methods in target tracking: A survey. IEEE Trans. Aerosp. Electron. Syst. 34, 103–123.

McHugh, M.L., 2012. The Chi-square test of independence. Biochem. Medica 23, 143–149.

McNeil, J.D., 1980. Electromagnetic Terrain Conductivity Measurement at Low Induction Numbers. Tech. note TN.

Melo, D.C., Régis, C.R.T., 2017. 1D laterally constrained inversion of 2D MT data. pp. 154–159.

Meola, C., Carlomagno, G.M., 2004. Recent advances in the use of infrared thermography. Meas. Sci. Technol. 15, 27.

Mertens, L., Persico, R., Matera, L., Lambot, S., 2016. Automated Detection of Reflection Hyperbolas in Complex GPR Images with No A Priori Knowledge on the Medium. IEEE Trans. Geosci. Remote Sens. 54, 580–596.

Metje, N., Atkins, P.R., Brennan, M.J., Chapman, D.N., Lim, H.M., Machell, J., Muggleton, J.M., Pennock, S., Ratcliffe, J., Redfern, M., 2007. Mapping the underworld–state-of-the-art review. Tunn. Undergr. Sp. Technol. 22, 568–586.

Metje, N., Chapman, D.N., Walton, R., Sadeghioon, A.M., Ward, M., 2012. Real time condition monitoring of buried water pipes. Tunn. Undergr. Sp. Technol. 28, 315–320.

Middleton, W.M., Valkenburg, M.E. van, 2002. Reference data for engineers, 9th ed, Reference data for engineers. Elsevier.

Milsom, J, Eriksen, A., 2011. Field Geophysics—fourth edition.

Milsom, John, Eriksen, A., 2011. Field geophysics, 4th ed, John Wiley & Sons Ltd. John Wiley & Sons Ltd.

Mix, P.E., 2005. Introduction to nondestructive testing: a training guide. John Wiley & Sons.

Moeller, J., Young, B., Ho, D.K.C., Anderson, D., 2021. Wire detection by GPR using the Hough transform. In: Detection and Sensing of Mines, Explosive Objects, and Obscured Targets XXVI. SPIE, pp. 32–43.

Moghadas, D., 2019. Probabilistic Inversion of Multiconfiguration Electromagnetic Induction Data Using Dimensionality Reduction Technique: A Numerical Study. Vadose Zo. J. 18, 1–16.

Moghadas, D., André, F., Saussez, S., Van Durmen, R., Van Leeuwen, C., Delvaux, B., Vereecken, H., Lambot, S., 2010. Soil moisture determination using ground penetrating radar and electromagnetic induction data in a ... In: GeoDarmstadt.

Mogharehabed, T., 2014. Experimental investigation of cast iron corrosion. University of Birmingham.

Moilanen, J., 2022. EMC Measurement Feasibility Study for Automated Test System.

Monteiro Santos, F.A., 2004. 1D laterally constrained inversion of 2D MT data. J. Appl. Geophys. 56, 123–134.

Monteiro Santos, F.A., El-Kaliouby, H.M., 2011. Quasi-2D inversion of DCR and TDEM data for shallow investigations. Geophysics 76.

Monteiro Santos, F.A., Triantafilis, J., Taylor, R.S., Holladay, S., Bruzgulis, K.E., 2010. Inversion of conductivity profiles from EM using full solution and a 1-D laterally constrained algorithm. J. Environ. Eng. Geophys. 15, 163–174.

Myung, I.J., 2003. Tutorial on maximum likelihood estimation. J. Math. Psychol. 47, 90–100.

Nadim, G., 2008. Clutter reduction and detection of landmine objects in ground penetrating radar data using likelihood method. 2008 3rd Int. Symp. Commun. Control. Signal Process. ISCCSP 2008 98–106.

Nelder, J.A., Mead, R., 1965. A simplex method for function minimization. Comput. J. 7, 308–313.

Ng, W., Chan, T.C.T., So, H.C., Ho, K.C., 2008. Particle filtering based approach for landmine detection using ground penetrating radar. IEEE Trans. Geosci. Remote Sens. 46, 3739–3755.

Nguyen, V.H., Pyun, J.Y., 2015. Location detection and tracking of moving targets by a 2D IR-UWB radar system. Sensors 15, 6740–6762.

Núñez-Nieto, X., Solla, M., Gómez-Pérez, P., Lorenzo, H., 2015. Signal-to-noise ratio dependence on ground penetrating radar antenna frequency in the field of landmine and UXO detection. Meas. J. Int. Meas. Confed. 73, 24–32.

Olhoeft, G.R., 2000. Maximizing the information return from ground penetrating radar. In: Journal of Applied Geophysics. Elsevier, pp. 175–187.

Oliveira, R.J., Caldeira, B., Teixidó, T., Borges, J.F., Bezzeghoud, M., 2022. Geophysical data fusion of ground-penetrating radar and magnetic datasets using 2D wavelet transform and singular value decomposition. Front. Earth Sci. 10, 1011999.

Onyszko, K., Fryśkowska-Skibniewska, A., 2021. A new methodology for the detection and extraction of hyperbolas in GPR images. Remote Sens. 13, 4892.

Özdemir, C., Demirci, Ş., Yiğit, E., Yilmaz, B., 2014. A review on migration methods in b-scan ground penetrating radar imaging. Math. Probl. Eng. 2014.

Pan, Q., Liu, D.J., Guo, Z.Y., Fang, H.F., Feng, M.Q., 2016. Magnetic anomaly inversion using magnetic dipole reconstruction based on the pipeline section segmentation method. J. Geophys. Eng. 13, 242–258.

Panda, S.L., Maiti, S., Sahoo, U.K., 2022. Subsurface Propagation Velocity Estimation

Methods in Ground-Penetrating Radar: A review. IEEE Geosci. Remote Sens. Mag. 10, 70–89.

Park, Sehwan, Kim, J., Jeong, S., Park, Seunghee, 2019. GPR Data-Based Computer Vision for the Detection of Material Buried Underground. In: Proceedings - 2019 3rd International Conference on Smart Grid and Smart Cities, ICSGSC 2019. Institute of Electrical and Electronics Engineers Inc., pp. 41–44.

Pasolli, E., Melgani, F., Donelli, M., 2009. Automatic analysis of GPR images: A pattern-recognition approach. IEEE Trans. Geosci. Remote Sens. 47, 2206–2217.

Pasternak, M., Kaczmarek, P., 2019. Continuous wave ground penetrating radars: state of the art. In: SPIE 11055, XII Conference on Reconnaissance and Electronic Warfare Systems. SPIE, p. 10.

Pennock, S.R., Abed, T.M., Curioni, G., Chapman, D.N., John, U.E., Jenks, C.H.J., 2014. Investigation of soil contamination by iron pipe corrosion and its influence on GPR detection. In: Proceedings of the 15th International Conference on Ground Penetrating Radar. IEEE, pp. 381–386.

Pennock, S.R., Chapman, D.N., Rogers, C.D.F., Royal, A.C.D., Naji, A., Redfern, M.A., 2010. Effects of iron pipe corrosion on GPR detection. In: Proceedings of the XIII Internarional Conference on Ground Penetrating Radar. IEEE, pp. 1–5.

Perrone, A., Lapenna, V., Piscitelli, S., 2014. Electrical resistivity tomography technique for landslide investigation: A review. Earth-Science Rev.

Persico, R., Soldovieri, F., Utsi, E., 2010. Microwave tomography for processing of GPR data at Ballachulish. J. Geophys. Eng. 7, 164–173.

Petersen, M.L., 2001. Use of Electromagnetic Induction Tools in Salinity Assessment/Appraisals in Eastern Colorado ERP1–ERP1.

Peyton, A.J., 2015. Electromagnetic induction tomography. In: Industrial Tomography: Systems and Applications. Woodhead Publishing, pp. 61–107.

Pilu, M., Fitzgibbon, A.W., Fisher, R.B., 2002. Ellipse-specific direct least-square fitting. In: Proceedings of 3rd IEEE International Conference on Image Processing. IEEE, pp. 599–602.

Prego, F.J., Solla, M., Puente, I., Arias, P., 2017. Efficient GPR data acquisition to detect underground pipes. NDT E Int. 91, 22–31.

Pridmore, D.F., Hohmann, G.W., Ward, S.H., Sill, W.R., 1981. An investigation of finite-element modeling for electrical and electromagnetic data in three dimensions. Geophysics 46, 1009–1024.

Pringle, J.K., Ruffell, A., Jervis, J.R., Donnelly, L., McKinley, J., Hansen, J., Morgan, R., Pirrie, D., Harrison, M., 2012. The use of geoscience methods for terrestrial forensic searches. Earth-Science Rev. 114, 108–123.

Pringle, J.K., Stimpson, I.G., Wisniewski, K.D., Heaton, V., Davenward, B., Mirosch, N., Spencer, F., Jervis, J.R., 2020. Geophysical monitoring of simulated homicide burials for forensic investigations. Sci. Rep. 10, 1–13.

Pringle, J.K., Stimpson, I.G., Wisniewski, K.D., Heaton, V., Davenward, B., Mirosch, N.,

Spencer, F., Jervis, J.R., 2021. Geophysical monitoring of simulated clandestine burials of murder victims to aid forensic investigators. Geol. Today 37, 63–65.

Qiu, H., Qiu, M., Lu, Z., Gerard, M., 2019. An efficient key distribution system for data fusion in V2X heterogeneous networks. Inf. Fusion 50, 212–220.

Radiodetection, 2018. The theory of buried cable and pipe location. Bristol.

Rajiv, K., Chandra, G.R., Rao, B.B., 2017. GPR Objects Hyperbola Region Feature Extraction.

Rao, K.R., Kim, D.N., Hwang, J.-J., 2010. Fast Fourier Transform - Algorithms and Applications. Signals and Communication Technology.

Reynolds, J.M., 2011. An Introduction to Applied and Environmental Geophysics, 2nd ed, European Space Agency, (Special Publication) ESA SP. Wiley.

Rizzo, P., 2010. Water and wastewater pipe nondestructive evaluation and health monitoring: A review. Adv. Civ. Eng. 2010.

Robbes, D., 2006. Highly sensitive magnetometers-a review. Sensors Actuators, A Phys. 129, 86–93.

Rodrigues, D., Barraca, N., Costa, A., Borges, J., Almeida, F., Fernandes, L., Moura, R., Madureira-Carvalho, Á., 2021. Multi-technique detection of buried inert explosive devices in urban context: Metal detection, magnetometer and ground-penetrating radar. In: Symposium on the Application of Geophysics to Engineering and Environmental Problems. Society of Exploration Geophysicists, pp. 218–219.

Rogers, C., Overton, C.G., Cohn, A., Pennock, S.R., C.H.J., J., Muggleton, J., Rustighi, E., Atkins, P., Foo, K.Y., Cross, J., Swingler, S., Chapman, D., Curioni, G., Royal, A.C.D., Metje, N., Parker, J., 2012. Mapping the Underworld.

Rosin, P.L., 2001. Unimodal thresholding. Pattern Recognit. 34, 2083–2096.

Roubal, M., 2018. Non-destructive testing of ground & structural integrity of underground assets. In: ISRM International Symposium 2000, IS 2000. OnePetro.

Saey, T., Van Meirvenne, M., De Smedt, P., Stichelbaut, B., Delefortrie, S., Baldwin, E., Gaffney, V., 2015. Combining EMI and GPR for non-invasive soil sensing at the stonehenge world heritage site: The reconstruction of a WW1 practice trench. Eur. J. Soil Sci. 66, 166–178.

Sagnard, F., Tarel, J.P., 2016. Template-matching based detection of hyperbolas in ground-penetrating radargrams for buried utilities. J. Geophys. Eng. 13, 491–504.

Samadzadegan, F., Toosi, A., Dadrass Javan, F., 2025. A critical review on multi-sensor and multi-platform remote sensing data fusion approaches: current status and prospects. Int. J. Remote Sens. 46, 1327–1402.

Samis, M., Davis, G.A., Laughton, D., Poulin, R., 2005. Valuing uncertain asset cash flows when there are no options: A real options approach. Resour. Policy 30, 285–298.

SanFilipo, B., 2000. Final report on passive and active low-frequency electromagnetic spectroscopy for airborne detection of underground facilities.

Santos, F.A.M., Triantafilis, J., Bruzgulis, K.E., Roe, J.A.E., 2010. Inversion of Multiconfiguration Electromagnetic (DUALEM-421) Profiling Data Using a One-Dimensional Laterally Constrained Algorithm. Vadose Zo. J. 9, 117.

Särkkä, S., 2013. Bayesian filtering and smoothing, 1st ed, Cambridge University Press. Cambridge University Press.

Sasaki, Y., 1989. Two-dimensional joint inversion of magnetotelluric and dipole data. Soc. Explor. Geophys. 54, 55–57.

Sasaki, Y., 2000. Full 3-D inversion of electromagnetic data on PC. J. Appl. Geophys. 46, 45–54.

Sbartaï, Z.M., Breysse, D., Larget, M., Balayssac, J.P., 2012. Combining NDT techniques for improved evaluation of concrete properties. Cem. Concr. Compos. 34, 725–733.

Schlinger, C.M., 1990. Magnetometer and gradiometer surveys for detection of underground storage tanks. Bull. - Assoc. Eng. Geol. 27, 37–50.

Schmidt, A., Chapman, D., Rogers, C., 2006. Physiochemcial changes in London Clay adjacent to cast iron pipes. Proc. Int. Assoc. Eng. Geol. Environ. Conf. 2006 12.

Scott, W.R., Kim, K., Larson, G.D., Gurbuz, A.C., McClellan, J.H., 2004. Combined seismic, radar, and induction sensor for landmine detection. J. Acoust. Soc. Am. 123, 3042–3042.

Shakmak, B., Al-Habaibeh, A., 2015. Detection of Water Leakage in Buried Pipes Using Infrared Technology.

Shihab, S., Al-Nuaimy, W., 2005. Radius estimation for cylindrical objects detected by ground penetrating radar. Subsurf. Sens. Technol. Appl. 6, 151–166.

Shihab, S., Al-Nuaimy, W., Huang, Y., Eriksen, A., 2002. Neural network target identifier based on statistical features of GPR signals. In: Koppenjan, S., Lee, H. (Eds.), Ninth International Conference on Ground Penetrating Radar. International Society for Optics and Photonics, pp. 135–138.

Simon, D., 2006. Optimal State Estimation: Kalman, H∞, and Nonlinear Approaches, Optimal State Estimation: Kalman, H∞, and Nonlinear Approaches.

Siu, K.L., Lai, W.W.L., 2019. A lab study of coupling effects of electromagnetic induction on underground utilities. J. Appl. Geophys. 164, 26–39.

Skartados, E., Kargakos, A., Tsiogas, E., Kostavelis, I., Giakoumis, D., Tzovaras, D., 2019. GPR antenna localization based on a-scans. Eur. Signal Process. Conf. 2019-Septe, 1–5.

Smith, R.E., Anderson, D.T., Ball, J.E., Zare, A., Alvey, B., 2017. Aggregation of Choquet integrals in GPR and EMI for handheld platform-based explosive hazard detection. In: Detection and Sensing of Mines, Explosive Objects, and Obscured Targets XXII. p. 1018217.

Smitha, N., Singh, V., 2016. Clutter reduction techniques of ground penetrating radar for detecting subsurface explosive objects. 2016 Int. Conf. Inf. Commun. Embed. Syst. ICICES 2016 1–8.

Solimene, R., Cuccaro, A., Dell'Aversano, A., Catapano, I., Soldovieri, F., 2014. Ground clutter removal in GPR surveys. IEEE J. Sel. Top. Appl. Earth Obs. Remote Sens. 7, 792–798.

Soltanova, D., Baranov, P., Baranova, V., Chudinova, A., 2015. Simulating magnetic field of a ferromagnetic pipe underwater in COMSOL Multiphysics. IOP Conf. Ser. Mater. Sci. Eng. 93, 012023.

Sophian, A., Tian, G., Fan, M., 2017. Pulsed eddy current non-destructive testing and evaluation: A review. Chinese J. Mech. Eng. 30, 500–514.

Srimuk, P., Boonpoonga, A., Kaemarungsi, K., Athikulwongse, K., Torrungrueng, D., Chantasen, N., 2023. Hyperbolic Pattern Detection in Ground Penetrating Radar Images Using Faster R-CNN. In: 2023 20th International Conference on Electrical Engineering/Electronics, Computer, Telecommunications and Information Technology, ECTI-CON 2023. Institute of Electrical and Electronics Engineers Inc.

Steinway, W.J., Duvoisin, H.A., Tomassi, M.S., Thomas, J.E., Betts, G., Morris, C., Kahn, B., Stern, P., Krywick, S., Johnson, K., Dennis, K., Simoneaux, W., Blood, B., 1998. Multi-sensor system for mine detection. In: IGARSS '98. Sensing and Managing the Environment. 1998 IEEE International Geoscience and Remote Sensing. Symposium Proceedings. (Cat. No.98CH36174). pp. 228–230 vol.1.

Sterling, R.L., Anspach, J.H., Allouche, E.N., Simicevic, J., Rogers, C.D.F., K. Weston, K.H., Weston, K.E., Hayes, K., 2009. Encouraging Innovation in Locating and Characterizing Underground Utilities, Encouraging Innovation in Locating and Characterizing Underground Utilities. National Academies Press.

Stölben, I.R.S., Bertelmann, J., Beltle, M., Tenbohlen, S., Bersch, C., Spanos, K., 2022. Investigation of Ground Impedances effecting EMC during Charging Operations of Electric Vehicles. In: 2022 International Symposium on Electromagnetic Compatibility– EMC Europe. IEEE, pp. 457–460.

Szymczyk, M., Szymczyk, P., 2013. Preprocessing of GPR data. ipc 18, 83–90.

Tabbagh, A., Panissod, C., 2000. 1D complete calculation for electrostatic soundings interpretation. Geophys. Prospect. 48, 511–520.

Terrasse, G., Nicolas, J.M., Trouve, E., Drouet, E., 2015. Application of the curvelet transform for pipe detection in GPR images. In: International Geoscience and Remote Sensing Symposium (IGARSS). pp. 4308–4311.

Thitimakorn, T., Kampananon, N., Jongjaiwanichkit, N., Kupongsak, S., 2016. Subsurface void detection under the road surface using ground penetrating radar (GPR), a case study in the Bangkok metropolitan area, Thailand. Int. J. Geo-Engineering 7, 1–9.

Torrione, P., Collins, L., 2006. Ground response tracking for improved landmine detection in ground penetrating radar data. In: International Geoscience and Remote Sensing Symposium (IGARSS). pp. 153–156.

Travassos, X.L., Avila, S.L., Ida, N., 2018. Artificial Neural Networks and Machine Learning techniques applied to Ground Penetrating Radar: A review.

Triantafilis, J., Santos, F.A.M., 2011. Hydrostratigraphic analysis of the Darling River valley (Australia) using electromagnetic induction data and a spatially constrained algorithm for quasi-three-dimensional electrical conductivity imaging. Hydrogeol. J. 19, 1053–1063.

Tso, C.H.M., Kuras, O., Wilkinson, P.B., Uhlemann, S., Chambers, J.E., Meldrum, P.I., Graham, J., Sherlock, E.F., Binley, A., 2017. Improved characterisation and modelling of measurement errors in electrical resistivity tomography (ERT) surveys. J. Appl. Geophys. 146, 103–119.

Tzanis, A., 2017. A versatile tuneable curvelet-like directional filter with application to fracture detection in two-dimensional GPR data. Signal Processing 132, 243–260.

USAG, 2019. 2019 Utility Strike Damages Report.

USAG, 2020. 2020 Utility Strike Damages Report.

Utsi, E.C., 2017. Ground Penetrating Radar-Theory and Practice, Journal of Chemical Information and Modeling. Joe Hayton.

Van de Vegte, J., 2002. Crash Course In Processing Digital Signal Processing. In: Fundamentals of Digital Signal Processing. Prentice-Hall International, pp. 1–28.

Van Kempen, L., Sahli, H., 2001a. Signal processing techniques for clutter parameters estimation and clutter removal in GPR data for landmine detection. IEEE Work. Stat. Signal Process. Proc. 158–161.

Van Kempen, L., Sahli, H., 2001b. Signal processing techniques for clutter parameters estimation and clutter removal in GPR data for landmine detection. In: IEEE Workshop on Statistical Signal Processing Proceedings. pp. 158–161.

Venkatasubramanian, V., Leung, H., 2006. An EM-IMM based abrupt change detector for landmine detection. Detect. Remediat. Technol. Mines Minelike Targets XI 6217, 62172Z.

Viezzoli, A., Christiansen, A.V., Auken, E., Sørensen, K., 2008. Quasi-3D modeling of airborne TEM data by spatially constrained inversion. Geophysics 73.

VIVAX, M., 2011. VM-480B User Handbook.

Vrugt, J.A., Ter Braak, C.J.F., 2011. DREAM (D): an adaptive Markov Chain Monte Carlo simulation algorithm to solve discrete, noncontinuous, and combinatorial posterior parameter estimation problems. Hydrol. Earth Syst. Sci. 15, 3701–3713.

Wahab, S., 2013. ASSESSING THE CONDITION OF BURIED PIPE USING GROUND University of Birmingham Research Archive.

Wai-Lok Lai, W., Kind, T., Kruschwitz, S., Wöstmann, J., Wiggenhauser, H., 2014. Spectral absorption of spatial and temporal ground penetrating radar signals by water in construction materials. NDT E Int. 67, 55–63.

Waite, J.W., Wellaratna, R., 2006. Centerline and depth locating method for non-metallic buried utility lines.

Waite, J.W., Wellaratna, R., 2010. Sensor fusion for model-based detection in pipe and cable locator systems.

Wang, J., Su, Y., 2013. Fast detection of GPR objects with cross correlation and hough transform. Prog. Electromagn. Res. C 38, 229–239.

Wang, M., Luo, H., Cheng, J.C.P., 2021. Towards an automated condition assessment framework of underground sewer pipes based on closed-circuit television (CCTV) images. Tunn. Undergr. Sp. Technol. 110, 103840.

Wang, M.L., Lynch, J.P., Sohn, H., 2014. Sensor Technologies for Civil Infrastructures, Sensor Technologies for Civil Infrastructures.

Wang, P., Goddard, K.F., Lewin, P.L., Swingler, S., Atkins, P., Foo, K.Y., 2011. Magnetic

Field Measurement to Detect and Locate Underground Power Cable.

Wang, P., Lewin, P., Goddard, K., Swingler, S., 2010. Design and testing of an induction coil for measuring the magnetic fields of underground power cables. In: Conference Record of IEEE International Symposium on Electrical Insulation.

Warren, C., Giannopoulos, A., Giannakis, I., 2016. gprMax: Open source software to simulate electromagnetic wave propagation for Ground Penetrating Radar. j 209.

Weisenseel, R.A., Karl, W.C., Castanon, D.A., Miller, E.L., Rappaport, C.M., DiMarzio, C.A., 1999. Statistical fusion of GPR and EMI data. In: Dubey, A.C., Harvey, J.F., Broach, J.T., Dugan, R.E. (Eds.), Detection and Remediation Technologies for Mines and Minelike Targets IV. SPIE, p. 1179.

Witten, A.J., 2014. Handbook of Geophysics and Archaeology, Taylor & Francis. Taylor & Francis.

Wong, P.T.W., Lai, W.W.L., Sato, M., 2016. Time-frequency spectral analysis of step frequency continuous wave and impulse ground penetrating radar. In: Proceedings of 2016 16th International Conference of Ground Penetrating Radar, GPR 2016. Institute of Electrical and Electronics Engineers Inc.

Wunderlich, T., Wilken, D., Majchczack, B.S., Segschneider, M., Rabbel, W., 2022. Hyperbola detection with RetinaNet and comparison of hyperbola fitting methods in GPR data from an archaeological site. Remote Sens. 14, 3665.

Xiao, J., Gao, Q., Ling, Y., Yan, J., Liu, B., 2021. Research on hyperbola detection and fitting in GPR B-scan image. In: 2021 International Conference on Communications, Information System and Computer Engineering (CISCE). IEEE, pp. 266–270.

Xie, X., Qin, H., Yu, C., Liu, L., 2013. An automatic recognition algorithm for GPR images of RC structure voids. J. Appl. Geophys. 99, 125–134.

Xu, X., Gao, W., Zhang, D., Wang, Y., 2018. Research on the Difference Detection Method Based on GPR Data. 2018 17th Int. Conf. Gr. Penetrating Radar, GPR 2018 1–4.

Yamaguchi, T., Mizutani, T., 2021. Detection and localization of manhole and joint covers in radar images by support vector machine and Hough transform. Autom. Constr. 126, 103651.

Yang, L., Liu, S., Liu, Y., Liang, S., Shi, X., 2019. A Novel De-Noising Method for GPR Signal. In: 2019 International Symposium on Antennas and Propagation, ISAP 2019 - Proceedings.

Youn, H.-S., Chen, C.-C., 2002. Automatic GPR target detection and clutter reduction using neural network. In: Koppenjan, S., Lee, H. (Eds.), Ninth International Conference on Ground Penetrating Radar. SPIE, pp. 579–582.

Zadhoush, H., 2021. Numerical modelling of ground penetrating radar for optimization of the time-zero adjustment and complex refractive index model. The University of Edinburgh.

Zajícová, K., Chuman, T., 2019. Application of ground penetrating radar methods in soil studies: A review. Geoderma.

Zare, E., Li, N., Khongnawang, T., Farzamian, M., Triantafilis, J., 2020. Identifying potential leakage zones in an irrigation supply channel by mapping soil properties using electromagnetic induction, inversion modelling and a support vector machine. Soil Syst. 4, 1–18.

Zhabitskaya, A., Sheshkus, A., Arlazarov, V.L., 2024. HoughToRadon transform: new neural network layer for features improvement in projection space. In: Sixteenth International Conference on Machine Vision (ICMV 2023). SPIE, pp. 297–304.

Zhang, X., Xue, F., Wang, Z., Wen, J., Guan, C., Wang, F., Han, L., Ying, N., 2021. A novel method of hyperbola recognition in ground penetrating radar (GPR) B-Scan image for tree roots detection. Forests 12, 1019.

Zhang, Y., Collins, L., Yu, H., Baum, C.E., Carin, L., 2003. Sensing of unexploded ordnance with magnetometer and induction data: Theory and signal processing. IEEE Trans. Geosci. Remote Sens. 41, 1005–1015.

Zhao, S., Al-Qadi, I., 2017. Pavement drainage pipe condition assessment by GPR image reconstruction using FDTD modeling. j 154.

Zheng, X., Liu, S., 2011. The design of broadband electromagnetic method system with GPS simultaneous localization. In: Lecture Notes in Electrical Engineering. Springer, Berlin, Heidelberg, pp. 611–617.

Zhou, X., Chen, H., Hao, T., 2019. Efficient detection of buried plastic pipes by combining GPR and electric field methods. IEEE Trans. Geosci. Remote Sens. 57, 3967–3979.

Zhou, X., Chen, H., Li, J., 2018. An Automatic GPR B-Scan Image Interpreting Model. IEEE Trans. Geosci. Remote Sens. 56, 3398–3412.

Zou, L., Wang, Y., Giannakis, I., Tosti, F., Alani, A.M., Sato, M., 2020. Mapping and assessment of tree roots using ground penetrating radar with low-cost GPS. Remote Sens. 12, 1300.

Zoubir, A.M., Chant, I.J., Brown, C.L., Barkat, B., Abeynayake, C., 2002. Signal processing techniques for landmine detection using impulse ground penetrating radar. IEEE Sens. J. 2, 41–51.

Zwinkels, J., 2015. Light, Electromagnetic Spectrum. In: Encyclopedia of Color Science and Technology. pp. 1–8.